\documentclass[11pt]{article}
\usepackage{fullpage}
\usepackage{amsmath}
\usepackage{amssymb}
\usepackage[dvips]{graphicx}
\usepackage{psfrag}

\def\lrbicross{{\blacktriangleright\!\!\!\triangleleft}}

\newcommand{\Ref}[1]{(\ref{#1})}

\renewcommand{\t}{\theta}

\newcommand{\cD}{{\mathcal D}}

\newcommand{\cV}{{\mathcal V}}
\newcommand{\real}{\mathbb{R}}
\newcommand{\comp}{\mathbb{C}}

\newcommand{\SU}{\mathrm{SU}}
\newcommand{\SL}{\mathrm{SL}}

\newcommand{\be}{\begin{equation}}
\newcommand{\ee}{\end{equation}}
\newcommand{\beq}{\begin{eqnarray}}
\newcommand{\eeq}{\end{eqnarray}}

\renewcommand{\det}{\mathrm{det}}

\newcommand{\sixjsym}[6]{
 \left\{\begin{array}{ccc}
#1 &  #2 & #3\\
#4& #5 & #6
\end{array}\right\}}

\newtheorem{theo}{Theorem}
\newtheorem{definition}{Definition}
\newtheorem{prop}{Proposition}

\newcommand{\proof}{\noindent\textit{Proof : --}\\ }

\begin{document}
\begin{center}
\LARGE{\bf 6J Symbols Duality Relations}
\bigskip
\bigskip

{\normalsize \sc L.Freidel${}^{a,b}$
\footnote{\tt lfreidel@perimeterinstitute.ca}, K.Noui${}^c$\footnote{\tt noui@phys.univ-tours.fr}, Ph.Roche${}^d$\footnote{\tt roche@lpta.univ-montp2.fr}
 \\[0.5cm]}
{\normalsize\it $^{a}$  PI, 35 King Street North, Ontario, Waterloo,Canada N2J 2W9
}\\[1mm]

{\normalsize\it $^b$ Laboratoire de Physique, ENS Lyon, 46 All\'ee d'Italie, Lyon 69007, France }\\[1mm]
{\normalsize\it $^c$ F\'ed\'eration Denis Poisson, LMPT UMR 6038, Parc de Grandmont, 37200 Tours, France}\\[1mm] 
{\normalsize\it $^d$  LPTA,
Universit\'e Montpellier II, 34095 Montpellier, France }\\[1mm]

\vspace*{1.5cm}

\large{\bf Abstract}
\end{center}\footnotesize
It is known that the Fourier transformation of the square of $(6j)$ symbols has a simple expression in the case of $su(2)$ and $U_q(su(2))$ when $q$ is a root of unit. The aim of the present work is to unravel the algebraic structure 
behind these identities. We show that  the  double crossproduct construction 
$H_1\bowtie H_2$ of two Hopf algebras and the bicrossproduct construction $ H_2^{*}\lrbicross H_1$ are the Hopf algebras structures behind these identities by analysing different examples. We study the case  where $D= H_1\bowtie H_2$ is equal to the group algebra of $ISU(2)$, $SL(2,\mathbb C)$ and where $D$ is a quantum double of a finite group, of $SU(2)$ and of $U_q(su(2))$ when $q$ is real.  
\normalsize

\noindent

\section{Introduction}
While studying the expectation values of observables in Turaev-Viro model,
J.W.Barrett \cite{Barrett} found and proved a remarkable and quite intriguing identity on quantum $(6j)$ symbols of $U_q(su(2))$ for $q$ root of unity
which reads:
 \be \label{dualityqroot}
  \left\{\begin{array}{ccc}
j_{\widehat{01}} & j_{\widehat{02}} & j_{\widehat{03}}
\\ j_{\widehat{23}}& j_{\widehat{13}} & j_{\widehat{12}}
\end{array}\right\}^2_q =
\left(\frac{2}{k}\right)^3 \sum_{(i_{IJ})\in D_{k}} \left\{\begin{array}{ccc}
i_{{01}} & i_{{02}} & i_{{03}}\\ i_{{23}}& i_{{13}} & i_{{12}}
\end{array}\right\}^2_q
\prod_{I<J}\sin \left( \frac{\pi}{k}(j_{IJ}+1)(i_{IJ}+1)\right)
\ee
 where $q=\exp(\frac{i\pi}{k}), k$ an integer $\geq 2$, $D_{k}=\{0,...,k-2\}^{\times 6},$ and
 $\widehat{IJ}=\{0,1,2,3\}\setminus \{IJ\}.$

This is an example of a self-duality formula for quantum $(6j)$ symbols: {\it the finite Fourier transform of the square of the function $(6j)$ is the square of the function $(6j)$ of the dual variables.}

While studying the asymptotics for large spins of classical $(6j)$ symbols of $su(2)$ \cite{FL6j},
the first author of the present work found and proved the following formula on volumes of flat tetrahedra.

Let $V(l_{IJ})$  be the volume of the flat euclidean
tetrahedron which is such that the length of the edge $(IJ)$ is
given by $l_{IJ}$. $V^{2}(l_{IJ})$ is a polynomial (known as the Cayley determinant) in the variables $l_{IJ}^2.$
We denote $V^*(l_{IJ})$ the volume of the dual tetrahedron i.e
$V^*(l_{IJ})=V(l_{\widehat{IJ}}).$

We denote $\cD_{\infty}$ the set of flat euclidean tetrahedra i.e
 $\cD_{\infty}$ is the subset of $l_{IJ} \in (\real^+)^6$
 satisfying the conditions  :
  $ l_{IJ} \leq l_{IK}+l_{KJ}$ for any triple
$(I,J,K)$ of distinct elements as well as $V^{2}(l_{IJ})\geq 0.$

 The inverse of the  volume satisfies the
remarkable self-duality property
\begin{equation}\label{dualityclassvol}
\frac{1}{ V^*(j _{IJ})}= \left(\frac{2}{
\pi}\right)^3\int_{\cD_{\infty}} \
\frac{\prod_{I<J}\sin(j_{IJ}l_{IJ}) dl_{IJ}}{ V(l_{IJ})}.
\end{equation}
The first duality formula (\ref{dualityqroot}) is purely algebraic  while the second one (\ref{dualityclassvol})
 involves geometrical objects.
It was shown in \cite{FL6j} that there are seemingly related identities
which have a mixed (algebraic and geometric) character relating
classical (6j) symbols and spherical geometrical objects, for instance:
\be \label{dualityclass1}
  \left\{\begin{array}{ccc}
j_{\widehat{01}} & j_{\widehat{02}} & j_{\widehat{03}}\\
 j_{\widehat{23}}& j_{\widehat{13}} & j_{\widehat{12}}
\end{array}\right\}^2 = \frac{2}{\pi^4}
\int_{\cD_{\pi}} \
\frac{\prod_{I<J}\sin\left((j_{IJ}+1)\theta_{IJ}\right)d\theta_{IJ}}{\sqrt{\det[\cos(\theta_{IJ})]}},
\end{equation}
where $\cD_{\pi}$ is  the set of spherical tetrahedra and $\theta_{IJ}$ are spherical
length of the edge $(IJ)$ i.e
 $\cD_{\pi}$ is the subset of $\theta_{IJ} \in [0,\pi]^6$
 satisfying the conditions  :
  $ \theta_{IJ} \leq \theta_{IK}+\theta_{KJ}$, $\t_{IJ} +\t_{IK}+\t_{KJ}\leq 2\pi$
   for any triple
$(I,J,K)$ of distinct elements as well as $\det[\cos(\theta_{IJ})]> 0$. The matrix
$\cos(\theta_{IJ})$ is known as the Gram matrix of the spherical tetrahedra.
Also by Fourier inversing the previous relation one obtains:
\be \label{dualityclass2}
\frac{\pi^2}{2^5}\frac{1}{\sqrt{\det[\cos(\theta_{IJ})]}}
=
\sum_{(j_{IJ})\in D_{\infty}} \left\{\begin{array}{ccc}
j_{\widehat{01}} & j_{\widehat{02}} & j_{\widehat{03}}\\ j_{\widehat{23}}& j_{\widehat{13}} & j_{\widehat{12}}
\end{array}\right\}^2
\prod_{I<J}\sin \left((j_{IJ}+1)\theta_{IJ}\right)
\ee
where $D_{\infty}=({\mathbb N})^{\times 6}.$

While studying cosmological deformation of spin foam models, the last authors of the present work discovered in \cite{NR} a relation between $(10j)$
symbols of simple principal representations of $U_q(so(3,1))$ and Fourier transform of product
of $(6j)$ symbols in the case where $q$ is real, which has striking similarities with a
duality transformation.
Note that these duality relations are playing a key role in the understanding of how particles and fields can be consistently coupled
to three dimensional gravity \cite{PRI,BarrettFeynman,PRIII,Noui}.

The aim of the present article is to unravel the algebraic structure lying behind these duality formulae. We will show that the natural algebraic interpretation of duality relations is expressed as a duality   between two Hopf algebras: the double cross product and the bicross product. Let  $(H_1, H_2)$ be two Hopf algebras and assume that $H_1$ $H_2$ is a matched pair in the sense of S.Majid \cite{Ma} so that one can define the double cross product $D=H_1\bowtie H_2.$ 
S.Majid has shown that the axioms defining the double cross product  
 $D=H_1\bowtie H_2$  imply the existence of another Hopf algebra denoted 
$\tilde{D}= H_2^{*}\lrbicross H_1$ which is coined bicrossproduct of $H_2^*$ and $H_1.$
One can define a notion of simple representation of $D$ (resp.$\tilde{D}$) as being a representation which admits a non zero invariant vector under $H_1.$
We will show for specific choices of $H_1,H_2$  that the duality relation for $D$ expresses that   the $(6j)$ of simple representations of $D$ are the Fourier transform of the $(6j)$ of simple representations of $\tilde{D}.$
We explain the different choices of $(H_1,H_2)$ we have made and we expect that the duality relation is more general.

Let us briefly present the first exemple: we consider $H$ a Poisson Lie group \cite{CP} and denote $H^{*}$ its associated dual Poisson Lie group. The Lie-Poisson double group  of $H$ \cite{CP} is  a Poisson Lie group 
${\cal D}(H)$ which contains $H$ and $H^*$ as sub Poisson-Lie groups of ${\cal D}(H).$
We can define a notion of simple representations of ${\cal D}(H)$ as being irreducible representations of  
${\cal D}(H)$ which admits a non zero invariant vector under the action of $H$.
The duality relation expresses a way to decompose certain  $(6j)$ symbols of simple representations of ${\cal D}(H)$ in term of  its Fourier modes.
 We analyze in detail the simplest case where $H=SU(2),$ in this case there is no multiplicities involved in the definition of $(6j)$ coefficients.

With the nul Poisson Lie group structure, we obtain $H^*=\mathbb{R}^3$ and 
${\cal D}(H)=ISU(2)$ and show that the simple $(6j)$ symbols are  proportional to the inverse of the volume $ V(l_{IJ})$. This leads to  the selfduality relation of the inverse
volume of flat tetrahedra.
In this case, we have  $H_1={\mathbb C}[SU(2)]$ and $H_2={\mathbb C}[\mathbb R^3].$ As a result 
$D= {\mathbb C}[SU(2)]\bowtie {\mathbb C}[\mathbb R^3]={\mathbb C}[ISU(2)],$ and $\tilde{D}=
{\mathbb C}[\mathbb R^3]^*\lrbicross  {\mathbb C}[SU(2)]={\mathbb C}[ISU(2)].$ Because $D$ and
 $\tilde{D}$ are isomorphic we obtain a self duality relation.

With the standard Poisson Lie group structure on $SU(2)$ we obtain that $H^*=AN(2)$ as well as ${\cal D}(H)=SL(2,\mathbb{C})$. The 
corresponding duality relation is the relation  (\ref{dualityslan}). In this case we have  $H_1={\mathbb C}[SU(2)]$ and $H_2={\mathbb C}[AN].$ As a result 
$D= {\mathbb C}[SU(2)]\bowtie {\mathbb C}[AN]={\mathbb C}[SL(2,{\mathbb C})],$ and $\tilde{D}=
{\mathbb C}[AN]^*\lrbicross  {\mathbb C}[SU(2)].$ Note that although $D$ is cocommutative, $\tilde{D}$ is not.  The duality relation which results is not a duality relation between $(6j)$ of simple representations of groups but rather a duality 
between $(6j)$ of simple representations of $SL(2,{\mathbb  C})$ and  $(6j)$ of simple representations  of a the Hopf algebra 
${\mathbb C}[AN]^*\lrbicross  {\mathbb C}[SU(2)].$

Then we focus on the very interesting of the quantum double $D(A)$ of a Hopf algebra $A.$
The quantum double of $A$ contains $A$ and $A^*$ as  Hopf subalgebras.
A simple representation of $D(A)$ is an irreducible representation which contains a non zero  invariant vector under the action of $A$. We show that the duality relation still express a way to decompose certain $(6j)$ symbols of simple representations in term of its Fourier modes. 
We analyze in detail the case where $A$ is successively the group algebra of a finite group, the compact group $SU(2)$ and the quantum groups $U_q(su(2))$  for $q$ real.
In these cases $H_1=A$ and $H_2=A^{*cop}$, as a result $D=D(A)$ and  $\tilde{D}=A_{op}\lrbicross A$ which can be shown to be isomorphic to $A^{op}\otimes A$ as a Hopf algebra. We therefore obtain a duality relation between $6j$ of simple representations of $D(A)$ and $6j$ of simple representations of  $A^{op}\otimes A.$

\section{Algebraic formulation of duality relations.}
In this section we first give a short review of evaluation of simple spin-networks in the case of a classical group and particularly in the case of a double of a Poisson Lie group $H$. 
We study the particular case where $H=SU(2)$: we define the different (classical) double structures, classify the simple representations, compute the
associated $(6j)$ symbols and proove duality relations.

Then, we give an analog study in the quantum case where the duality relation is a relation between $(6j)$ symbols of the quantum double of a 
Hopf algebra $A$ and the $(6j)$ symbols of the Hopf algebra itself.

Finally, we observe that these results are in fact exemples of a more general case which involves double crossproduct and bicrossproduct algebra 
structures.

\subsection{General results on a classical  group}
Let $G$ be a Lie group and $H$ be a subgroup of $G.$
Let us consider an irreducible unitary representation of $G$, denoted $\pi,$ and we will denote
 $\langle,\rangle$ the hermitian form  associated to this unitary representation acting on the vector space
 $V_{\pi}.$ $\pi$ is said to be a {\it simple} representation of $G$ if it  admits a non zero $H$-invariant vector $\omega$
(note that in the mathematic litterature these representations are called  {\it class one}  representations.)
In the case that we will study the $H$-invariant vectors are unique up to a multiplicative constant.
If $\pi_1,...,\pi_n$ are simple representations, we denote 
$\iota^s:\bigotimes_{j=1}^n V_{\pi_j}\rightarrow \mathbb{C}$ the {\it symmetric} intertwiner defined by 
\begin{equation}\label{Symint}
\iota^s(\otimes_{j=1}^n v_j)=\int_{G}\prod_{j=1}^n\langle \omega_j,\pi_j(g)v_j\rangle dg,\end{equation}
where $dg$ is the Haar measure on $G$ (it is defined up to a scalar and we will choose the normalized one for compact $G$.)

Simple spin-networks $\Gamma$  are spin-networks whose edges $l$  are colored with simple representations $\pi_l$ and 
vertices $I$ with the symmetric  intertwiner.
From the $H$-invariance of $\omega_j$ the 
  symmetric  intertwiner can be defined  as an integral over the coset $G/H$.  As a result the evaluation of  a simple spin-network reduces to an integration over $(G/H)^{\times p}$ where $p$ is the number of vertices of the simple spin-network. The general construction of symmetric intertwiner and the representation of the evaluation of a simple spin network in term of Feynman graph has been done in \cite{FK}. Note that in the case where the coset space $G/H$ is non compact, the evaluation of the spin network   needs a regularization which is usually taken by selecting any vertex $I_0$ and integrating
 only over the remaining  $p-1$ vertices.  One obtains a function on $G/H$  which is constant and whose value on any point  gives the evaluation $ev(\Gamma)$ of the spin network \cite{Baezbarrett}.

Given a simple representation $\pi$, the   propagator $K_\pi(x,y)$ is a function on $G\times G$ given by the following matrix element:
\begin{eqnarray}\label{Kpi}
K_\pi(x,y) \; = \; \langle\omega \vert \; \pi(xy^{-1}) \; \omega\rangle\;.
\end{eqnarray}
The symmetric kernel is given by $K_\pi(x)=K_\pi(x,e).$ As a result the evaluation of a simple spin network is given by:
\begin{equation}
ev(\Gamma)=\int_{(G/H)^{p-1}}\prod_{I,J} K_{\pi_{l_{IJ}}}(x_I,x_J) \prod_{I\not=I_0} dx_I.
\end{equation}

We will be particularly interested in the case where $H$ is a Poisson Lie group and $G={\cal D}
(H)$ is the double of $H.$
Let $H$ be a connected simply-connected Poisson Lie group in the sense of Drinfeld \cite{CP}. The Lie algebra ${\mathfrak h}$ of $H$ is a Lie bialgebra. As a result ${\mathfrak h}^*$ is also a Lie bialgebra which defines uniquely a connected simply-connected group denoted $H^*$ and called the dual group of $H.$
One can define on the vector space $\mathfrak{h}\oplus {\mathfrak h}^*$ a  structure of Lie bialgebra denoted ${\cal D}({\mathfrak h})$, it is such that 
$\mathfrak{h}, {\mathfrak h}^{*op}$ are sub-Lie bialgebras ( ${\mathfrak h}^{*op}$ is the dual Lie bialgebra of $\mathfrak{h}$ with opposite cobracket). As a result this Lie bialgebra defines an unique connected and simply connected Poisson Lie group denoted ${\cal D}(H)$ which contains $H$ and $H^*$ as sub Lie Poisson groups.
We shall assume that as a manifold we have ${\cal D}(H)=H\times H^*=G.$ Note that $H$ is generally not normal in $G$ but nevertheless the homogeneous space $G/H$ can be endowed with the  structure of group $H^*$. 
The Poisson structure on $H^*$ is degenerate and we denote ${\cal S}$ the set of symplectic leaves of $H^*.$ 
A very important notion is the dressing action of $H$ on $H^*$: let $g,g'\in H,\gamma,\gamma'\in H^*$, the product in ${\cal D}(H)$ satisfies
\begin{equation}
g\gamma.g'\gamma'=g ({\gamma\triangleright g'})({\gamma\triangleleft g'})\gamma'
\end{equation} 
where ${\gamma\triangleright g'}\in H$ and ${\gamma\triangleleft g'} \in H^*.$
The function $H\times H^*\rightarrow H^*, (g,\gamma)\mapsto {\gamma\triangleleft g}$ is  called the dressing action of $H$ on
 $H^*.$
 The symplectic leaves of $H^*$ can be shown to be   the orbits of the dressing action of $H$ on $H^*$. The  propagator $K_{\pi}$ is a function on the homogeneous space $G/H=H^*.$ From the relation
 $\gamma g=(\gamma\triangleright g) (\gamma\triangleleft g) $ and the $H$-invariance of $\omega$ one obtains that  the function $K_{\pi}$ is constant on each orbit of the dressing action i.e $K_{\pi}$ is a constant function on each symplectic leaf. If $l \in {\cal S}$ we denote $K_{\pi l}$ the value of $K_{\pi}$ on this leaf.
The function $K_{\pi}$ can therefore be written as
\begin{equation}
K_{\pi}(x)=\int_{{\cal S}}K_{\pi l}  1_{l}(x)dl
\end{equation}
where $dl$ is the measure on the set of symplectic leaves and $1_l$ is the characteristic function of $l.$

\begin{figure}
\psfrag{0}{$1$}
\psfrag{3}{$4$}
\psfrag{1}{$2$}
\psfrag{2}{$3$}
\psfrag{A1}{$\pi_{12}$}
\psfrag{A2}{$\pi_{23}$}
\psfrag{A3}{$\pi_{34}$}
\psfrag{A4}{$\pi_{14}$}
\psfrag{A5}{$\pi_{24}$}
\psfrag{A6}{$\pi_{13}$}
\centering
\includegraphics[scale=0.7]{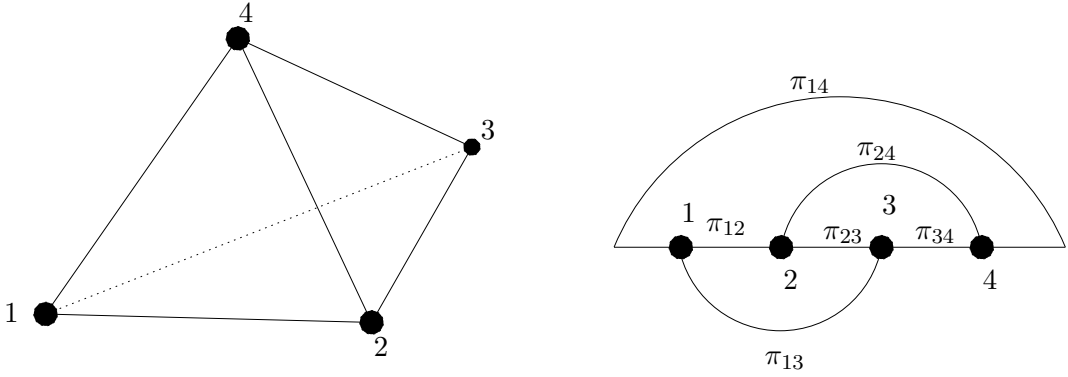}
\caption{Tetrahedron spin-network $T$. The set of vertices are ordered and we label them by $I \in \{1,2,3,4\}$; edges are labelled by a couple $(IJ)$ colored with representations $\pi_{IJ}$. }
\label{tetrahedron}
\end{figure}

The evaluation of the tetrahedral spin network (figure \ref{tetrahedron}) associated to simple irreps $\pi_{IJ}$ of $G$ gives:
\begin{equation}
T(\pi_{IJ})=\int_{(H^*)^3}\prod_{I<J}K_{\pi_{IJ}}(x_I x_J^{-1})dx_1 dx_2 dx_3
\end{equation}
this function being constant in $x_4.$
Using the expression of $K_{\pi}$ we obtain:
\begin{equation}
T(\pi_{IJ})=\int_{{\cal S}^6}\int_{(H^*)^3}\prod_{I<J}K_{\pi_{IJ}l_{IJ}} 1_{l_{IJ}}(x_I x_J^{-1})dx_1 dx_2 dx_3.
\end{equation}
Let us define $\tilde{T}(l_{IJ})=\int_{(H^*)^3}\prod_{I<J}1_{l_{IJ}}{(x_I x_J^{-1})}dx_1 dx_2 dx_3,$ the evaluation of the tetrahedral spin network satisfies:
\begin{equation}
T(\pi_{IJ})=\int_{{\cal S}^6}\prod_{I<J}K_{\pi_{IJ}l_{IJ}} \tilde{T}(l_{IJ}) \; dl_{IJ}\; .\label{generalclassicalduality}
\end{equation}

Before studying this relation in different cases, let us remark some points.
First, it is clear that, from the very definition of $K_\pi$ (\ref{Kpi})
and from the definition of the symmetric intertwiner (\ref{Symint}), $T(\pi_{IJ})$ is (up to a normalisation we have to compute) a $(6j)$ symbol.
Thus, this relation underlines a duality between $(6j)$ symbols and some others functions $\tilde{T}(l_{IJ})$ that we are going to interpret 
as $(6j)$ symbols as well in the following.

The rest of this section is devoted to the study of this relation in the case where $H$ is $SU(2)$ by evaluating precisely the functions 
$T(\pi_{IJ})$ and $ \tilde{T}(l_{IJ})$.

\subsubsection{The example of the euclidean group $ISU(2)$.}
We first assume that $H$ is endowed with its structure of nul Poisson bracket. In that case the Drinfeld Double of $H$ is simply the cotangent bundle $T^*(H)=H\ltimes {\mathfrak h}^*.$
We will only  study the case where $H=SU(2)$ and in that case one can identify $T^*(SU(2))$ with the euclidean group $G=ISU(2).$
 It is defined as the semi-direct product $ISU(2)= SU(2)\ltimes \mathbb R^3 $ and  any of its elements is written as  $(u,\vec{a})$ where $\vec{a}$ is a vector in $\mathbb R^3$ and $u$ an element of $SU(2)$. We have 
$(u,\vec{a})(v,\vec{b})=(uv, \Lambda(v^{-1})\vec{a}+\vec{b})$ where $\Lambda$ denotes the vectorial representation of 
$ SU(2).$

Conjugacy classes of $SU(2)$ are labelled by an angle $\theta \in [0,2\pi]$ and we will denote by $h_\theta$ a representative of the conjugacy class.

Unitary Irreducible representations of $ISU(2)$ are labelled by a couple $(m,s) \in \mathbb R^+ \times \mathbb N$ and consist of the vector spaces defined by ${\cal V}_{m,s}=\{\phi:SU(2) \rightarrow \mathbb C \vert \phi(x h_\theta)=e^{i\theta s} \phi(x)\}$. The action of $(u,\vec{a})$ on this space is given by:
\begin{eqnarray}
[\rho^m_s(u,\vec{a})\phi](x) \; = \; e^{i\vec{a}\cdot \vec{p}(u^{-1}x)} \; \phi(u^{-1}x)\;,
\end{eqnarray}
where  $\vec{p}(x)=\Lambda(x)\vec{M}$ and $\vec{M}=(m,0,0)$. The vector space ${\cal V}_{m,s}$ is  endowed with the Hilbert structure defined from the normalized $SU(2)$ Haar measure $dx$. The representation $\rho^m_s$ is an irreducible  unitary representation.

\medskip

Simple representations with respect to the sub-group $SU(2)$ are those which admits a non zero $SU(2)$-invariant vector. It is immediate to show that simple representations are such that $s=0$ and ${\cal V}_{m,0} $ is therefore identified with functions on $SU(2)/U(1)=S^2.$ 
In this case the $SU(2)$ invariant vector is the constant function $\omega=1$ on the sphere. The  propagator is the function   on $\mathbb R^3$  given by:
\begin{eqnarray}
K_m(\vec{a}) \; = \; \int_{SU(2)} dx \; \overline{\omega(x)}\; e^{i \vec{a}\cdot \Lambda(x)\vec{M}}\;\omega(x)\;.
\end{eqnarray}
A  straightforward calculation shows that:
\begin{eqnarray}\label{Kernel of ISU(2)}
K_m(\vec{a}) \; = \; \frac{1}{4\pi}\int_{{\mathbb R}^3} \frac{d^3\vec{k}}{k^2} \; \delta(k-a) \; e^{i\vec{k}\cdot \vec{M}}\; = \;\frac{1}{4\pi} \int_{S^2} d^2\vec{n}\; e^{ia\vec{n}\cdot \vec{M}}\;
\end{eqnarray}
where we have denoted $k=\vert\vert\vec{k}\vert\vert$ and $a=\vert\vert\vec{a}\vert\vert$.

We will now compute the datas of the general classical duality relation (\ref{generalclassicalduality}).
The dressing action of $SU(2)$ on $\mathbb{R}^3$ is the vectorial action and the symplectic leaves are the 2-spheres.
The integral  $K_m(\vec{a})$ is straighforwardly done and one gets:  
\begin{eqnarray}
K_m(\vec{a}) \; = \; \frac{\sin(ma)}{ma} \;.
\end{eqnarray}

Let us now evaluate the tetrahedron simple spin-network $T$ colored with representations $(m_{IJ})$. To do so, we associate a  propagator to each edge of the tetrahedron and we obtain a function $F_T(m_{IJ})$ on $({\mathbb R}^{3})^{\times 4}$. Each argument of the function is associated to a vertex of the tetrahedron. Because  $\mathbb R^3$ is non compact we regularize this integral by choosing one vertex $I_0=4$ and  integrate the function $F_T(m_{IJ})$ over all the vertices but $I_0$: the obtained result is a function proportional to the identity in $\mathbb R^3$ whose proportionnality coefficient is denoted $T(m_{IJ})$. By definition, $T(m_{IJ})$ is the evaluation of the simple spin-network and one can show that it is independent of the choice of the chosen  vertex. Therefore, we have:
\begin{eqnarray}
T(m_{IJ}) \; = \; \int \prod_{I<J} K_{m_{IJ}}(\vec{a}_{IJ})\;  \prod_{I\not=I_0}d\vec{a}_{I}\; \text{with} \; \; \vec{a}_{IJ}=\vec{a}_I - \vec{a}_J
\end{eqnarray}
that we can write of the form (\ref{generalclassicalduality}) as follows:
\begin{eqnarray}
T(m_{IJ}) \; = \; \int_{}\prod_{I<J} dk_{IJ} \frac{\sin(m_{IJ}k_{IJ})}{m_{IJ}k_{IJ}}\tilde{T}(k_{IJ}) \;\;\;\;\; \text{with} \;\;\;\;\;
\tilde{T}(k_{IJ})=\int_{}\prod_{I\not=I_0}d\vec{a}_{I}\delta(\vert a_{IJ}\vert-k_{IJ}) \;.
\end{eqnarray}
To go further into the calculations,  one can use at this point the proposition $(6)$ of \cite{FL6j} which states that:
\begin{equation}
\prod_{I\not=I_0}\frac{d^3 \vec{a}_I}{2\pi}=\frac{1}{3\pi V(a_{IJ})}\prod_{I<J}a_{IJ}d a_{IJ},
\end{equation}
from which we get 
 $\tilde{T}(k_{IJ})=\frac{8 \pi^2}{3} \frac{\prod_{I<J}k_{IJ}}{V(k_{IJ})}.$
This proves that:
\begin{equation}
T(m_{IJ})=\frac{8\pi^2}{3}\int_{}\prod_{I<J} dk_{IJ} \frac{\sin(m_{IJ}k_{IJ})}{m_{IJ}}\frac{1}{V(k_{IJ})}.
\end{equation}

Moreover, one can compute explicitely the value of $T(m_{IJ})$. A direct proof is provided in the proof of Theorem 3 of  \cite{FL6j} and one obtains: 
\begin{equation}T(m_{IJ})=\frac{\pi^5}{3}
\frac{1}{ V^*(m_{IJ})\prod_{I<J}m_{IJ}}
\end{equation}
 i.e one relates directly $T(m_{IJ})$ to the volume of the dual flat tetrahedron. From this relation one gets 
the duality relation (\ref{dualityclassvol}).

It remains to express $T(m_{IJ})$ in term of the $(6j)$ coefficients of the simple representations $\rho^{m_{IJ}}_{0}$ of $ISU(2).$
The reader is invited to read the appendix  of the present work for normalisation issues.
We have the relation 
\begin{equation}
T(m_{IJ})=2\pi^2\prod_{K=1}^4 \sqrt{\Theta_K(m_{IJ})}
 \sixjsym{m_{12}}{m_{13}}{m_{14}}{m_{34}}{m_{24}}{m_{23}}\;,
\end{equation}
where the functions $\Theta_K(m_{IJ})$ have been computed in the appendix and is given by:
\begin{eqnarray}
\Theta(m_1,m_2,m_3)&=&\frac{4\pi}{m_1m_2m_3} \int_0^{+\infty} da \; \frac{\sin(m_1 a)\sin(m_2 a) \sin(m_3 a)}{a} \\
&=&\frac{\pi^2}{m_1 m_2 m_3} \; Y(m_1,m_2,m_3),\end{eqnarray}
where we have introduced the function $Y(m_1,m_2,m_3)$ which takes the value $1$ or $0$ if $m_1,m_2,m_3$ satisfy triangular inequalities or not.

 Therefore the expression of  the $(6j)$ symbol $\left\{\begin{array}{ccc}
m_{{12}} & m_{{13}} & m_{{14}}\\
 m_{{34}}& m_{{24}} & m_{{23}}
\end{array}\right\}$ of simple representations of $ISU(2)$ is  given by:
\begin{eqnarray}
\left\{\begin{array}{ccc}
m_{{12}} & m_{{13}} & m_{{14}}\\
 m_{{34}}& m_{{24}} & m_{{23}}
\end{array}\right\} \; = \; \frac{1}{6\pi} \frac{1}{V^*(m_{IJ})}\;.
\end{eqnarray}
The classical duality relation (\ref{dualityclassvol}) can  therefore also  be expressed as an auto-duality relation between 
the $(6j)$ symbols of simple representations  of $ISU(2).$

\subsubsection{The example of the Lorentz group $SO(3,1)$.}
We now consider the group $H=SU(2)$ with its non zero standard Lie-Poisson structure.
It is well known that in this case the dual Poisson group of $SU(2)$ is the Borel subgroup  $B=AN$ of  $SL(2,\mathbb{C})$. The double of $SU(2)$ is the group $G=SL(2,\mathbb{C})$ universal cover of 
$SO(3,1).$
 The principal series of unitary representations of $G$ is labelled by
a couple $(\rho,s)$ where $\rho$ is a real number and $s$ an integer.
We will focus on the simple representations for which $s=0$, these are the  representations admiting a normalised
$SU(2)$ invariant vector $\omega$.
We define the  propagator to be the function on $G^{\times 2} $
\be K_\rho(x,y)=\langle\omega,\pi_\rho(xy^{-1})\omega\rangle. \ee
This propagator defines a function on   the hyperboloid ${\cal H}_3=SL(2,\mathbb{C})/SU(2)$.

$\SL(2,\comp)$ naturally acts as a group of isometry of Minkowski space 
${\mathbb R}^{1,3}.$
The carrier space of simple representations $\cV_\rho$
is the space of homogeneous function of degree $-1+i\rho$
on the forward light-cone  ${\cal C}=\{\xi\in {\mathbb R}^{1,3},  \xi^2 =0, \xi_0>0\}$\cite{Baezbarrett}:
\be
\cV_\rho = \{\phi:{\cal C}\rightarrow 
{\mathbb C},\ \phi(\lambda \xi) = \lambda^{-1+i\rho} \phi(\xi),
 \xi\in {\cal C},\lambda>0 \}.
\ee
The inner product is given by the integral
\be
\langle\phi,\psi \rangle = \int_{S^2} \bar{\phi}(\xi)\psi(\xi) d\xi
\ee
over the 2-sphere of null vectors satisfying $\xi_0=1$.
The action of $\SL(2,\comp)$ on $\cV_\rho$ is induced by its action on the light cone
$ g\cdot \phi(\xi) = \phi(g^{-1} \xi)$.
The $\SU(2)$ invariant vector $\omega(\xi)$ is the unit function on $S^2$.
The  propagator is a function on  ${\cal H}_3^{\times 2}$ given by
\be\label{prop}
K_\rho(x,y)= \int_{S^2} \bar{\omega}(\xi) |x\cdot\xi|^{-1-i\rho}|y\cdot\xi|^{-1+i\rho} \omega(\xi)d\xi
= \frac{\sin(\rho r)}{\rho \sinh r},
\ee
where $\cdot$ denote the Minkowski scalar product and $r=r(x,y)$ is the hyperbolic
distance between  $x$ and $y$.

In order to evaluate simple spin-network, a preliminary work consists in computing the measure on ${\cal H}_3$. For that, we proceed as follows.
The polar decomposition of the group $SL(2,\mathbb C)$ says that for any $g\in G$
there exists a unique $(h,k) \in\mathbb H_+ \times  SU(2) $ such that $g=hk.$ 
This implies that that the 3-dimensional hyperbolic space ${\cal H}_3$
is isometric to ${\mathbb H}_+ $ the set  of positive  hermitian two by two matrices with unit determinant:
\be
{\cal H}_3 \cong {\mathbb H}_+ \; = \;
\{h \in {\mathbb H} \; \vert \; \det h = 1 \; \text{and} \; \text{tr}\;h>0 \},
h= \left(\begin{array}{cc} x_0+x_3 & x_1+ix_2 \\ x_1-ix_2 & x_0 -x_3 \end{array}\right).
\ee
The geometric action of $SL(2,{\mathbb C})$
is given by the action $h \mapsto ghg^\dagger$ on hermitian matrices.

Another decomposition of $SL(2,\mathbb C)$, the Iwasawa decomposition, provides another description of the
hyperboloid which says that there exists a unique $(k,a,n)$ in $SU(2) \times A \times N$
such that $g=ank$,
where $A$ is the group of diagonal positive matrices of determinant 1 and $N$ is the nilpotent group of
upper triangular matrices with diagonal elements equal to 1.
The Iwasawa decomposition implies that the hyperboloid can also be identified as:
\begin{eqnarray}\label{isoH3AN2}
{\cal H}_3 \cong\; AN=
 \{  \left( \begin{array}{cc} a & n \\ 0 & a^{-1} \end{array} \right)
 \; \text{with} \; a \in \mathbb R^{*+} \; \text{and} \; n \in \mathbb C \}.\end{eqnarray}
Expressing the polar decomposition in term of the Iwasawa one is straightforward and amounts to solve the equation $h^2=an (an)^{\dagger}.$
${\cal H}_3$ can also be described as the submanifold defined by $-x_0^2+x_1^2+x_2^2+x_3^2=1$. If we introduce $r>0$ by $x_0=\cosh r,$ and $\vec{x}=(\sinh r) \vec{n}$ with $ \vec{n}\in S^2,$ the volume element on the hyperboloid is $d^3x=\sinh^2 r dr d^2 \vec{n}.$

We can now apply the relation  (\ref{generalclassicalduality}) in the present case.
Consider one tetrahedra  whose edges $(IJ)$ are colored by simple 
representations $(\rho_{IJ})$ of $\SL(2,\comp)$.
We have 
\begin{equation}
T(\rho_{IJ}) \; =
\int_{}\prod_{I<J} dl_{IJ} \frac{\sin(\rho_{IJ}l_{IJ})}{\rho_{IJ}
\sinh(l_{IJ})}\tilde{T}(l_{IJ}),
\end{equation}
where $\tilde{T}(l_{IJ})=\int_{{{\cal H}_3^{\times 3}}}
 \prod_{I=1}^3 d^3x_I 
 \delta( r(x_I,x_J)-l_{IJ}).$

One can proceed analogously to the $ISU(2)$ and one obtains the following theorem.

\begin{theo}[Duality relation for $\SL(2,\comp)$]
The evaluation of the tetrahedral spin network is given by:
\begin{equation}
T(\rho_{IJ}) \; =16\pi^2
\int_{\cD_\infty^H}\prod_{I<J} dl_{IJ} \frac{\sin(\rho_{IJ}l_{IJ})}{\rho_{IJ}}
 \frac{1}{\sqrt{|\det[-\cosh(l_{IJ})]|}},\label{dualityslan}
\end{equation}
where we have denoted $\cD_\infty^H$ the set of hyperbolic tetrahedra.

If we denote $\left\{\begin{array}{ccc}
 \rho_{{12}}& \rho_{{13}} & \rho_{{14}}\\
\rho_{{34}} & \rho_{{24}} & \rho_{{23}}
\end{array}\right\}$ the $(6j)$ symbols of the simple representions $\rho_{IJ}$ of $SL(2,\mathbb{C})$ the following identity holds:
\begin{equation}
\left\{\begin{array}{ccc}
 \rho_{{12}}& \rho_{{13}} & \rho_{{14}}\\
\rho_{{34}} & \rho_{{24}} & \rho_{{23}}
\end{array}\right\}\prod_{K=1}^4 \sqrt{\Theta_K(\rho_{IJ})})=8\int_{\cD_\infty^H}\prod_{I<J} dl_{IJ} \frac{\sin(\rho_{IJ}l_{IJ})}{\rho_{IJ}}
\frac{1}{\sqrt{|\det[-\cosh(l_{IJ})]|}}.
\end{equation}

where the function $\Theta$ is defined in the appendix and given by: 
\begin{eqnarray}
\Theta(\rho_1,\rho_2,\rho_3)
= \frac{4\pi}{\rho_1\rho_2\rho_3} \int_0^\infty \frac{\sin \rho_1 r \,\sin \rho_2 r\, \sin \rho_3r}{\sinh r}dr
 =\frac{\pi^2}{2\rho_1\rho_2\rho_3}\left( \tanh(\frac{\pi}{2}(\rho_1+\rho_2-\rho_3))\right. \\
 +\left.\tanh(\frac{\pi}{2}(\rho_1-\rho_2+\rho_3))
+\tanh(\frac{\pi}{2}(-\rho_1+\rho_2+\rho_3))-\tanh(\frac{\pi}{2}(\rho_1+\rho_2+\rho_3))\right).
\end{eqnarray}
\end{theo}
\proof
The second part of the theorem follows immediately from the first and from the expression of $\Theta$ given in the appendix.
The first part of the theorem amounts to compute   $\tilde{T}(l_{IJ})$. This is analogous to the analysis performed in  \cite{FL6j} which 
 expresses the measure in terms of the invariant lengths and 
 we obtain:
\be\label{meashyp}
\prod_{I=1}^3\frac{d^3 x_I}{2\pi}
= \frac{2}{\pi} \frac{\prod_{IJ}\sinh l_{IJ}}{\sqrt{|\det[-\cosh l_{IJ}]|}}dl_{IJ}.
\ee
From this result and the expression of the propagator \Ref{prop} we have
\be \tilde{T}(l_{IJ})=16\pi^2 \prod_{I<J}\sinh (l_{IJ})\frac{1}{\sqrt{|\det[-\cosh l_{IJ}]|}}.
\ee
$\Box$

We now study the quantum case and begin with a reminder of results on the quantum double of a Hopf algebra.

\subsection{General Results on the Quantum Double}

Let $A$ be a Hopf algebra, $D(A),$ the quantum double of $A$, is a quasitriangular Hopf algebra containing $A$ as a sub-Hopf algebra and has different equivalent definitions that we present here.

If $A$ is a Hopf algebra we denote $m$ the multiplication, $\Delta$ the coproduct and $\sigma:A^{\otimes 2}\rightarrow A^{\otimes 2}$ the permutation operator. From $A$ we can construct different Hopf-algebras as follows: $A_{op}$ is the Hopf algebra with multiplication $m_{op}=m\circ \sigma$ and coproduct $\Delta; $ $A^{cop}$ is the Hopf algebra with multiplication $m$ and coproduct
 $\Delta^{cop}=\sigma\circ \Delta.$
As usual $A^*$ is naturally endowed with a Hopf algebra structure and we denote by $(e_i)$ a basis of $A$ , $(e^i)$ the  dual basis.

A right and left invariant Haar measure on a Hopf algebra $A$ is a linear form $h:A\rightarrow {\mathbb C}$ satisfying:
\begin{equation}
(h\otimes id)\Delta(a)=h(a) 1\;\; \text{and}\;\;\;\; (id\otimes h)\Delta(a)=h(a) 1.
\end{equation}

Let $\pi,\pi'$ be representations of $A$ acting on $V$ and $V',$  let  $\phi\in V^*$ and $v'\in V'$. If $h$ is a Haar measure on $A^*$  we formally define an interwiner $\iota[\pi,\pi']_{\phi,v'}:V\rightarrow V'$ by
\begin{equation}
\iota[\pi,\pi']_{\phi,v'}(v)=\sum_{i,j}\pi'(S(e_i))(v') \langle \phi, \pi(e_j)(v)\rangle h(e^ie^j).
\end{equation}
Note that this is formal because this series needs to be shown to be convergent in some precise sense in the ``non-compact'' case (see for example $U_q(so(3,1))$).

Let us now present one  definition of the quantum double (the Drinfeld double):

\begin{definition}[The Drinfeld double]
$D(A)$ is  the  Hopf algebra defined as:
\begin{enumerate}
\item $D(A)=A\otimes A^{* cop}$ as a coalgebra
\item the algebra law is given by:
\begin{equation}(x\otimes \xi)(y\otimes\eta)=\sum_{(y),(\xi)}
xy_{(2)}\otimes \xi_{(2)}\eta S^{-1}y_{(1)}(\xi_{(1)})
 y_{(3)}(\xi_{(3)})
\end{equation}
where we have used Sweedler notation:
\begin{equation}
\Delta(x)=\sum_{(x)}x_{(1)}\otimes x_{(2)},\;\;
\Delta_{ A^{* cop}}(\xi)=\sum_{(\xi)}\xi_{(1)}\otimes \xi_{(2)}.
\end{equation}\end{enumerate}
$D(A)$ is quasitriangular and  the $R$ matrix is given by $R=\sum_{i}e_i\otimes 1\otimes 1\otimes e^i$.
\end{definition}
$D(A)$ being a Hopf algebra, its dual Hopf algebra $D(A)^*$ has a simple description that we recall here:
\begin{enumerate}
\item $D(A)^{*}=A^*\otimes A_{op}$ as an algebra
\item the coproduct on $D(A)^{*}$ is twist equivalent to the coproduct of $A^*\otimes A_{op}$ where the twist $F$ is given by $F^{-1}=\sum_{i}e_i\otimes e^i.$
\end{enumerate}

We recall here the general results that are known on the representation theory of  $D(A)$ and $D(A)^*.$ First, the algebra isomorphism $D(A)^*\simeq  A^*\otimes A_{op}$ implies that a finite dimensional irreducible representation of $D(A)^*$ is a  tensor product of an irrep of $A$ with an irrep of $A^*.$ In the case of $D(A)$, if $A$ is quasitriangular and factorizable
\footnote{For example $U_q(su(2))$ is factorizable whereas $F(SU(2))$ and 
${\mathbb C}[SU(2)]$ are not.} then  $D(A)$ is isomorphic as an algebra to $A\otimes A$. As a result, in that case, finite dimensional irreducible representations are tensor product of two finite dimensional irreps of A.

Let $\rho$ be a representation of $D(A)$, because $A\otimes 1$ is a subalgebra of $D(A),$ $\rho_{\vert A}$ is a representation of $A$.

\begin{definition}[Simple representations of a quantum double]
We will define $\rho$ to be a simple representation acting on $V$, if $\rho$ is an irreducible representation of $D(A)$ and if:

1. it  decomposes as a direct sum of irreducible representations of $A$

2. the trivial representation appears in this decomposition with multiplicity one.
We will denote by $\omega\in V$ a unit A-invariant vector, i.e $\forall a\in A, \rho(a)\omega=\epsilon(a)\omega$ where $\epsilon$ is the counit.

3. the space of linear form invariant under the contragredient representation $\rho^*$ is one dimensional and not contained into $(\mathbb{C}\omega)^{\perp}.$
As a result we will denote $\phi\in V^*$ such that
 $ \forall a\in A, {}^t\rho(a)\phi=\epsilon(a)\phi,$ with $\phi(\omega)=1.$

Remark that in the case of unitary representation the condition $(3)$ is implied by $(2)$ and in that case $\phi=\langle \omega\vert.$
\end{definition}

Note that if $\rho$ is a simple representation then the value of the square of the ribbon element in this representation is equal to one.
Let $R=\sum_i a_i\otimes b_i$, we define $u=\sum_i S(b_i)a_i$. By definition the ribbon element $v$ is defined as $v^2=uS(u)=S(u)u.$
As a result we obtain that $v^2=\sum_{i,j}a_jS(b_j)S(b_i)a_i.$
The expression of the $R$ matrix in the double is such that $a_i=e_i\otimes 1\in A\otimes 1$. Therefore if we evaluate $v^2$ in a simple representation we obtain
\begin{equation}
\langle \phi,v^2 \omega \rangle=\langle \phi,\sum_{i,j}a_jS(b_j)S(b_i)a_i\omega \rangle=\langle \phi,\epsilon(a_j)\sum_{i,j}S(b_j)S(b_i)\epsilon(a_i)\omega \rangle=1
\end{equation}
where we have used  $(\epsilon\otimes id)(R)=1.$

Simple representations are the basic ingredients in the general definition of
 the {\it symmetric} intertwiner.

\begin{definition}[Symmetric Intertwiner]\label{symmetricintertwiner}
We can define a general notion of symmetric intertwiner  as follows \cite{NR,BR2}.
We denote $(x^M)$ a basis of $D(A)$ and $(x_M)$ the dual basis.
Let $\rho_i, \rho_j'$ be   simple representations of $D(A)$ acting on $V_i, V_j'.$ We define the symmetric intertwiner of $D(A)$ as being the intertwiner $\iota^s[\rho,\rho']=\iota[\rho,\rho']_{\phi,\omega}$ with $\rho=\otimes_i \rho_i$, $\rho'=\otimes_j \rho_j'$, $\phi=\otimes_i \phi_i$ and $\omega=\otimes_j \omega_j,$ i.e
\begin{equation}
\iota^s[\rho,\rho']=\sum_{M,N}(\bigotimes_{j=1}^p \rho_j')(\Delta^{(p)}(S(x^M)))\vert \omega\rangle \langle \phi\vert( \bigotimes_{i=1}^n \rho_i)(\Delta^{(n)}(x^N) )
h(x_M x_N).
\end{equation}
\end{definition}

A simple spin-network is associated to a graph whose edges are colored with simple representations of $D(A)$ and vertices with symmetric  intertwiners. The evaluation of any simple spin-network reduces to a multi-integration of the  propagator defined as follows.

\begin{definition}[Symmetric Propagator]
Given a simple representation $\rho$, we denote the symmetric kernel as being the element $K_{\rho}\in (D(A))^{*}$ defined by $K_{\rho}=
\langle \phi\vert \rho(.)\vert\omega\rangle.$
The  propagator is the element $(id\otimes S)K_{\rho}\in (D(A))^{*\otimes 2}$ that we will conveniently denote $K_{\rho}(x,y).$
\end{definition}

We now give some examples where the representation theory of the quantum double is completely understood and we will construct explicitely the  propagator. We will show that the symmetric kernel is dual (in a sense we will precise) to the characters of unitary representations of the underlying Hopf algebra $A$. To be more precise, the  propagator can be expressed as the Fourier transform of these characters.  When one evaluates a simple spin-network associated to the tetrahedron, one gets a duality relations between $(6j)$ symbols of simple representations of $D(A)$ and $(6j)^2$ symbols of finite dimensional  representations of $A$.

We will start with the case where $A$ is a finite group $H$: even if this case has no particular physical interest, it is very interesting because it is a very nice and simple illustration of duality relations. Furthermore, it is a very general example which can be extended easily to the compact group case and can be illuminating in other cases. Then, we continue with the example of $A=SU(2)$ which is of particular interest for three dimensional euclidean quantum gravity. Then, we end up with the case of $A=U_q(su(2))$ for $q$ real and $q$ root of unity.

\subsubsection{The finite group case}

Let $H$ be a finite group, we denote by $F(H)$ and $\mathbb C[H]$ respectively the algebra of functions on $H$ and the group algebra endowed with the usual Hopf algebra structures. In the sequel, we will denote by $(\delta_g)_{g\in H}$ and $(x)_{x \in H}$ their respective basis.

\medskip

Let us start by presenting the Hopf algebra $D(H)$ defined as the quantum double of $\mathbb C[H]$. Note that this definition can be straightforwardly extended to the case where $H$  is a  compact Lie group \cite{KM} but beware that this algebraic object has not to be confused with ${\cal D}(H)$ the double of the Lie Poisson group $H$ if $H$ is a Lie Poisson group. As a coalgebra we have $D(\mathbb{C}[H])=\mathbb{C}[H]\otimes F(H)^{cop};$ thus a basis of $D(H)$ is $(x\otimes \delta_g)_{x,g\in H}$. The Hopf algebra structure in this basis is easily obtained and is given by:
\begin{eqnarray}
&&(x\otimes \delta_g).(y\otimes \delta_h)=xy\otimes\delta_h \delta_{h,y^{-1}gy}\\
&&\Delta(x\otimes \delta_g)=\sum_{g_1,g_2 \in H, g=g_1 g_2}(x\otimes \delta_{g_2})\otimes (x\otimes \delta_{g_1}).
\end{eqnarray}
The action of the antipode on the basis is $S(x\otimes \delta_g)=x^{-1}\otimes \delta_{x^{-1}g^{-1}x}.$

Note that one can also find in the litterature \cite{DPR} that the quantum double of a finite group is in fact defined as $D(F(H))$, the quantum double of the Hopf algebra $F(H)$. As a coalgebra we have $D(F(H))=F(H)\otimes \mathbb{C}[H]$. A basis of $D(H)$ is $(\delta_g \otimes x)_{x,g}$ and the structure of the Hopf algebra in this basis is given by:
\begin{eqnarray}
&&(\delta_g\otimes x).(\delta_h\otimes y)=\delta_{g,xhx^{-1}}\delta_g\otimes xy,\\
&&\Delta(\delta_g\otimes x)=\sum_{g_1,g_2\in H, g_1 g_2=g }(\delta_{g_1}\otimes x)\otimes
 (\delta_{g_2}\otimes x).
\end{eqnarray}
In this article, we have  prefered to work with the quantum double of $\mathbb{C}[H]$ as being the definition of  $D(H).$ This is completely equivalent because  $D(\mathbb{C}[H])$ and $D(F(H))^{cop}$ are isomorphic as Hopf algebras: an explicit isomorphism  $\psi$ being  given by $\psi(x\otimes \delta_g)=\delta_{xgx^{-1}}\otimes x.$

For  the sequel, let us consider the dual of $D(H)$. It is a Hopf algebra and its structure expressed in the  dual basis $(\delta_x\otimes g)_{g,x\in H}$ is  given by:
\begin{eqnarray}
&&( \delta_x\otimes g).( \delta_y\otimes k)=\delta_{x,y} (\delta_x\otimes
 kg),\\
&&\Delta(\delta_x\otimes g)=\sum_{x_1,x_2\in G, x_1 x_2=x}(\delta_{x_1}\otimes x_2gx_2^{-1})\otimes (\delta_{x_2}\otimes g ).
\end{eqnarray}
The antipode reads $S( \delta_x\otimes g)=\delta_{x^{-1}}\otimes xg^{-1}x^{-1}.$ $D(H)^{*}$ admits a normalized right and left invariant Haar measure  $h:D(\mathbb{C}[H])^{*}\rightarrow {\mathbb C}$ given by:
\begin{equation}
h(\delta_{x}\otimes g)=\frac{1}{\vert H\vert}\delta_{g,e}.
\end{equation}

\medskip

Now, let us present general results on the representation theory of $D(H)$. First, we recall that irreducible representations of $D(F(H))$ have been classified in \cite{DPR}; they are
denoted $\rho_{\pi}^{\theta}$ and labelled by a couple $(\theta,\pi)$ where $\theta$ is a conjugacy class of $H$ and $\pi$ an irreducible representation of the centralizer of any element of $\theta.$ To describe completely these representations, we will write  $\theta=\{g_1,...,g_r\}$ where $r=\vert \theta \vert$, denote  $N_\theta$  the centralizer of $g_1$, and define $g_k=x_k g_1 x_k^{-1}$ where $x_k$ is a section of $H/N_\theta.$ A basis of the representation  $\rho_{\pi}^{\theta}$ is denoted $\vert x_j,v_i\rangle,$ where $j=1,...,\vert \theta\vert$ and $(v_i)$ denotes a basis of the representation $\pi$. Thus, the representation $\rho_{\pi}^{\theta}$ is defined as:
\begin{equation}\label{representation of finite group}
 \rho_{\pi}^{\theta}(x\otimes \delta_g)\vert x_j,e_i\rangle=
\delta_{g,g_j}\vert x_k,\pi(h) e_i\rangle
\end{equation}
where $x x_j=x_k h$ with $h\in N_\theta.$ Equivalently, if we define  $\vert g_j,e_i\rangle=\vert x_j,e_i\rangle$, we have:
\begin{equation}
 \rho_{\pi}^{\theta}(x\otimes \delta_g)\vert g_j,e_i\rangle=
\delta_{g,g_j}\vert xg_jx^{-1},\pi(h) e_i\rangle.
\end{equation}
Simple representations are defined as possessing a non zero  $H$-invariant vector.

\begin{prop}[Simple representations of $D(H)$]
A representation  $\rho_{\pi}^{\theta}$ is simple if and only if $\pi$ is the trivial representation. In that case, the representation admits an unique normalized $H$-invariant vector $\omega$ given by:
\begin{equation}
\omega=\frac{1}{\sqrt{\vert \theta \vert }}\sum_{j=1}^{\vert \theta\vert }\vert x_j\rangle=
\frac{1}{\sqrt{\vert \theta \vert }}\sum_{g\in \theta}\vert g\rangle.\label{omegaG}
\end{equation}
\end{prop}

\proof
From the inclusion ${\mathbb C}[H]\hookrightarrow D(H)$,  $\rho_{\pi}^{\theta}$ defines a representation of $H$, therefore the  multiplicity $m$ of the trivial representation of $H$ in $\rho_{\pi}^{\theta}$ is given by:
\begin{eqnarray}
m=\frac{1}{\vert H\vert}\sum_{x\in H}\text{tr}(\rho_{\pi}^{\theta}( x\otimes 1))=\frac{1}{\vert H\vert}\sum_{x,g\in H}\text{tr}(\rho_{\pi}^{\theta}(x\otimes \delta_g)).
\end{eqnarray}
In \cite{DPR} it has been shown that:
\[
\text{tr}(\rho_{\pi}^{\theta}(x\otimes \delta_g))=
\left\{
\begin{array} {l}  0 \quad\text{if}\quad g\notin \theta \quad  \;\; \text{or} \;\;[g,x]\not= e\\
\text{tr}\;( \pi(h)) \text{where}\; xx_j=x_jh, g=x_j g_1 x_j^{-1}, h\in N_\theta.
\end{array}
\right.
\]
As a result, the multiplicity of the trivial representation reduces to:
\begin{eqnarray}
&&m=\frac{1}{\vert H\vert}
\sum_{x,g\in H, g\in \theta, [g,x]=e}\text{tr}\rho_{\pi}^{\theta}(x\otimes \delta_g)
\overline{\text{tr}\rho_{0}^{\theta}(x\otimes \delta_g)}\\
&&=\frac{1}{\vert H\vert}
\sum_{x,g\in H}\text{tr}\rho_{\pi}^{\theta}(x\otimes\delta_g)
\overline{\text{tr}\rho_{0}^{\theta}(x\otimes\delta_g)}=\delta_{\pi,0}.
\end{eqnarray}
Therefore, a representation $\rho_{\pi}^{\theta}$ is simple if and only if $\pi$ is the trivial representation. In that case, one can easily show that $\omega$ defined as in (\ref{omegaG}) is $H$-invariant.
$\Box$

\medskip

The symmetric intertwiner $\iota^s$ is obtained from  definition (\ref{symmetricintertwiner}) and can be computed explicitely using the Hopf algebra structure on $D(H)$. The symmetric propagator is easily computed and is given in the following proposition.
\begin{prop}[Symmetric Propagator of $D(H)$] Given a conjugacy class $\theta$, the symmetric  kernel $K_\theta$ is an element of $D(H)^{*}$ whose explicit expression reads:
\begin{equation}\label{Kernel of finite group}
K_{\theta}=\frac{1}{\vert \theta\vert}1\otimes \sum_{g\in \theta}g=
\frac{1}{\vert \theta\vert}1\otimes \theta \;\;\;\;\; \text{with} \;\; \theta= \sum_{g\in \theta} g.
\end{equation}
Therefore, $K_\theta$ can be viewed as an  element of $\mathbb C[H] \subset D(H)^{*}$, which is \, in fact, a consequence of the $H-invariance$ of the symmetric  intertwiner. $K_\theta$ can be expanded in ``Fourier modes'' as follows:
\begin{eqnarray}\label{Fourier transform propagator}
K_\theta = \frac{1}{\vert H \vert} \sum_j \overline{\chi_j(\theta)} \sum_{g\in H} \chi_j(g) g \;.
\end{eqnarray}
In that expression, $j$ labels unitary representations of $H$; $\chi_j$ denotes the character of the $j$ representation of $H$  and $\chi_j(\theta) = \chi_j(g_1)$ where $g_1$ is any representative of the conjugacy class $\theta$.

The  propagator is defined as $K_\theta(x,y) \equiv (id\otimes S)(\Delta(K_\theta))$, is an element of $({\mathbb C}[H])^{\otimes 2}$ and reads:
\begin{equation}
K_\theta(x,y)=\frac{1}{\vert \theta\vert}\sum_{g\in \theta} (1\otimes g)\otimes(1\otimes g^{-1})\;.\end{equation}
\end{prop}

\proof
Given a simple representation labelled by the conjugacy class $\theta$, the symmetric kernel is defined as $K_\theta =<\omega \vert \rho^\theta(\cdot) \vert \omega> \in D(H)^*$. Therefore, one can write the kernel in the basis of $D(H)^*$ as follows:
\begin{eqnarray}
K_\theta \; = \; \sum_{x,g} <\omega \vert \rho^\theta(x \otimes \delta_g) \vert \omega> \delta_x \otimes g \;.
\end{eqnarray}
Using the expression of the representation (\ref{representation of finite group}), one sees immediately that the kernel is given by the expression (\ref{Kernel of finite group}) and therefore is an element of $\mathbb C[H] \subset D(H)^*$. In an equivalent way, one can write it as:
\begin{eqnarray}
K_\theta \; = \; \frac{1}{\vert \theta\vert} 1 \otimes \theta \; = \; 
\frac{1}{\vert \theta \vert} 1 \otimes \sum_{g\in H} 1_\theta(g) \; g \;,\label{propagatorfinitegroup}
\end{eqnarray}
where $1_\theta \in F(H)$ is the characteristic function in the conjugacy class $\theta$. This function is obviously invariant under the adjoint action of the group and  can be expanded in terms of characters of unitary representations of $H$ as follows:
\begin{eqnarray}
1_\theta(g) \; = \; \frac{\vert \theta \vert}{\vert H \vert} \sum_j \overline{\chi_j(\theta)} \; \chi_j(g) \;.
\end{eqnarray}
As a result, one sees that the kernel is in fact the Fourier transform of the characters of the group $H$. The expression of the propagator is a direct consequence of the expression of the antipode on $D(H)^*$. $\Box$

\medskip

Now, all the ingredients are present to prove duality relations between $(6j)$ symbols of simple representations of $D(H)$ with $(6j)^2$ of unitary irreducible representations of $H$.
For that purpose, we need to evaluate simple spin-networks using the techniques already mentionned before.
However, we now have to care about the ordering in the multiplications of the different propagators for these functions do not commute
anymore. The order of multiplication is in fact fixed by the definition of the symmetric intertwiner at each vertex.

\begin{theo}[Duality in the case of $D(H)$ with $H$ a finite group]\label{Duality finite group}
Let us consider the simple spin-network associated to the tetrahedron graph $T$ colored with simple representations $(A_{IJ})$ of $D(H)$ (figure \ref{tetrahedron}). The evaluation of this spin-network that we will denote $T(\theta_{IJ})$ is a function of the six simple representations $\rho_0^{\theta_{IJ}}$ that is given by:
\begin{eqnarray}\label{duality in case of finite group}
T(\theta_{IJ}) \; = \; \frac{1}{\vert H\vert^3} \; \sum_{j_{IJ} \in Irr_H^{\times 6}} \left( \prod_{I<J} \overline{\chi_{j_{IJ}}(\theta_{IJ})}\right)  \left\{\begin{array}{ccc}
j_{{12}} & j_{{13}} & j_{{14}}\\
 j_{{34}}& j_{{24}} & j_{{23}}
\end{array}\right\}^2\;,
\end{eqnarray}
where we have denoted by $Irr_H$ the set of inequivalent unitary irreducible representations of $H$ and we have denoted 
\begin{equation}
 \left\{\begin{array}{ccc}
j_{{12}} & j_{{13}} & j_{{14}}\\
 j_{{34}}& j_{{24}} & j_{{23}}
\end{array}\right\}^2=\frac{1}{\vert H \vert^4} \sum_{(g_{I})\in H^{\times 4}} \prod_{I<J} \chi_{j_{IJ}}(g_{I} g_{J}^{-1}) \;.
\end{equation}

\end{theo}

\proof
The spin-network (figure \ref{tetrahedron}) is evaluated by integrating the following function $F_T(\theta_{IJ}) \in D(H)^{*\otimes 4}$:
\begin{eqnarray}\label{Tetrahedron function}
F_T(\theta_{IJ}) & = & K_{\theta_{14}(1)} K_{\theta_{12}(1)} K_{\theta_{13}(1)} \; \otimes \; S( K_{\theta_{12}(2)}) K_{\theta_{24}(1)} K_{\theta_{23}(1)} \nonumber \\
&\otimes & S(K_{\theta_{13}(2)}) S(K_{\theta_{23}(2)}) K_{\theta_{34}(1)} \; \otimes \; S(K_{\theta_{34}(2)} K_{\theta_{24}(2)} K_{\theta_{14}(2)})\;.
\end{eqnarray}
We have used the Sweedler notation $\Delta K_\theta = K_{\theta(1)} \otimes K_{\theta(2)}.$ If one uses the expression (\ref{Fourier transform propagator}), one shows immediately that the previous function can be expressed as the Fourier transform of  $\tilde{F}_T(j_{IJ}) \in \mathbb C[H]^{\otimes 4}$:
\begin{eqnarray}
F_T(\theta_{IJ}) & = & \sum_{j_{IJ}} \left( \prod_{I<J} \overline{\chi_{j_{IJ}}(\theta_{IJ})}\right) \tilde{F}_T(j_{IJ}) \;\;\; \text{with}\\
\tilde{F}_T(j_{IJ}) & = & \frac{1}{\vert H\vert^6} \sum_{g_{IJ}} \left( \prod_{I<J} \chi_{j_{IJ}}(g_{IJ})\right) g_{14}g_{12}g_{13}\otimes g_{12}^{-1} g_{24} g_{23} \otimes g_{13}^{-1} g_{23}^{-1} g_{34} \otimes g_{14}^{-1} g_{24}^{-1} g_{34}^{-1}\;.\nonumber
\end{eqnarray}
The next step consists on integrating this function on each argument using the fact that $\forall g \in \mathbb C[H] \subset D(H)^*, h(g)=\delta_e(g)$ where $h$ is the Haar measure on $\mathbb C[H]$. After performing the integrations and imposing  the delta functions, we obtain  that:
\begin{eqnarray}
h^{\otimes 4} (\tilde{F}_T(j_{IJ})) \; = \; \frac{1}{\vert H \vert^7} \sum_{g_{I}} \prod_{I<J} \chi_{j_{IJ}}(g_{I} g_{J}^{-1}) \;.
\end{eqnarray}
The theorem \ref{Duality finite group} follows immediately. $\Box$

\medskip

We gave here a general proof of this duality relation for any finite group. Nicely, this duality relation can be easily extended in the case of a  compact group $H$.  We will only study the case $H=SU(2)$ in the sequel. We will also show that this duality applies in the case of $U_q(sl(2))$ when $q$ is real;
the case where $q$ is a root of unity was considered by Barrett in (\ref{Barrett}).

\subsubsection{The compact group case $SU(2)$}
Let $H$ be the compact group $SU(2)$, we denote as in the finite group case by $F(SU(2))$ and $\mathbb C[SU(2)]$ respectively the algebra of  functions on $SU(2)$ and the group algebra endowed with the well-known following Hopf algebra structures.

The algebra structure of $\mathbb C[SU(2)]$ is naturally induced by the group structure of $SU(2)$ and the co-algebra structure is completely defined by:
\begin{eqnarray}
\forall \; g \in SU(2),\;\; \Delta(g)=g\otimes g,\;\; S(g)=g^{-1},\;\; \text{and}\; \epsilon(g)=\delta_e(g)\;.
\end{eqnarray}
There exists a right and left invariant Haar measure on $\mathbb C[SU(2)]$  defined by $h: \mathbb C[SU(2)] \rightarrow \mathbb C$ and $h(g)=\delta_e(g)$.

The algebra $F(SU(2))$ is commutative; the co-algebra structure of $F(SU(2))$ is obtained by duality with $\mathbb C[SU(2)]$ and reads:
\begin{eqnarray}
\forall \; f\in F(SU(2)),\;\;\forall x,y\in H,\; \Delta(f)(x,y) =f(xy),\;\; S(f)(x)=f(x^{-1})\;\; \text{and}\; \epsilon(f)=f(e)\;.
\end{eqnarray}

We denote as usual  by $dg$ the normalized  Haar measure on $SU(2).$ 

\medskip

Now, we briefly review the definition and elementary properties of
the quantum double   $D(SU(2))$ defined as the quantum double of
$\mathbb C[SU(2)]$. Note that one can also find in the litterature
$D(SU(2))$ defined as the quantum double of $F(SU(2))$ but we prefer to adopt the other  but equivalent definition here. As a
co-algebra, $D(SU(2))=\mathbb C[SU(2)] \otimes F(SU(2))^{cop}$ and
we will denote by $(x\otimes f)$ any of its elements. The Hopf
algebra structure is given by:
\begin{eqnarray}\label{Hopfstruct}
\forall \; f,g \in F(SU(2)) \;,\; x,y \in SU(2),&&\!\!\!\!\! (x\otimes f)(y \otimes g)=xy\otimes 
(f\circ Ad_y) g, \\
&& \!\!\!\!\! \Delta(x\otimes f)(a,b) = f(ba) \; x\otimes x \;.
\end{eqnarray}
The action of the antipode is given by $S(x\otimes f)(a)=x^{-1}\otimes f(x^{-1}a^{-1} x)$ for any $a \in SU(2)$.

Let us consider $D(SU(2))^*$ the dual Hopf algebra, its  structure is  expressed on any dual element of the form $(f\otimes x)$ by:
\begin{eqnarray}
(f \otimes x) (g \otimes y) \; = \; fg \otimes yx \\
\Delta(f\otimes x)(a,b) \; = \; f(ab)\; bxb^{-1}\otimes x \;.
\end{eqnarray}
The antipode reads $S(f\otimes x)(a)=f(a^{-1})\otimes ax^{-1}a^{-1}$. $D(SU(2))^*$ admits a right and left Haar measure defined as the tensor product of the Haar measure on $\mathbb C[SU(2)]$ and the Haar measure on $F(SU(2))$.

\medskip

Unitary irreducible representations of $D(SU(2))$ have been studied and are classified by a couple $(\theta,s)$: $\theta \in [0,2\pi[$
labels a conjugacy class of $SU(2)$ whose  representative is
chosen to be $h_\theta$; $s\in \mathbb Z$ is an
integer which labels irreducible representations of the
centralizer $C_\theta$ of the element $h_\theta$. Note that
$C_\theta=U(1)$ for $\theta \neq 0, 2\pi$; otherwise $C_\theta$ is
the group $G=SU(2)$ itself. In the generic case $\theta\neq 0$,
the vector space of a representation $(\theta,s)$ consists of the
subspace of functions on $SU(2)$ defined by ${\cal
V}_{\theta,s}=\{\varphi:SU(2) \rightarrow \mathbb C \vert \;
\forall \; a\in SU(2),\; \varphi(ah_\theta)=e^{is\theta}
\varphi(a)\}$. The representation of any element $(x\otimes f) \in
D(SU(2))$ on the above vector space reads:
\begin{eqnarray}\label{representation of D(SU(2))}
[\rho^\theta_s(x\otimes f)\psi](a) \; = \; f(ah_{\theta} a^{-1})\psi(x^{-1}a)\;.
\end{eqnarray}
The vector space ${\cal V}_{\theta,s}$ inherits a natural Hilbert space structure from the $L^2$ structure on $F(SU(2))$.

\begin{prop}[Simple representations and symmetric propagator of D(SU(2))]\label{Proposition compact group}

Simple representations of $D(SU(2))$   are of the form $(\theta,0)$. Any simple representation admits as unique normalized $SU(2)$-invariant vector the function $\omega=1$.

Given a simple representation labeled by $\theta$, the symmetric kernel $K_\theta$ is an element of $F(SU(2))^\star \subset D(SU(2))^\star$ given by:
\begin{eqnarray}
K_\theta \; : \; F(SU(2)) \longrightarrow \mathbb C \;\;,\;\;\;\; f \longmapsto K_\theta(f)=\int dg\; f(gh_\theta g^{-1}) \;.
\end{eqnarray}
 $K_\theta$ can also be viewed as an element of the Hopf algebra ${\mathbb C}[SU(2)] \subset D(SU(2))^*$ and can be expanded in Fourier modes as follows:
\begin{eqnarray}\label{Kernel of SU(2)}
K_\theta=1\otimes \int dg \; \delta_\theta(g)\;  g \; = 
1\otimes \; \sum_j \chi_j(h_\theta) \int dg\; \chi_j(g) g \;,
\end{eqnarray}
where $\delta_\theta$ is the characteristic function  of the conjugacy class $C_\theta.$
This expression is the continuous analog of formula (\ref{propagatorfinitegroup}).
The symmetric  propagator is defined as $K_\theta(x,y)=(id\otimes S)\Delta(K_\theta)$, is an element of ${\mathbb C}[SU(2)]^{\otimes 2}$ and reads:
\begin{eqnarray}
K_\theta(x,y) \; = \; \sum_j \chi_j(h_\theta) \int dg\; \chi_j(g) \; (1\otimes g)\otimes(1\otimes g^{-1})\;.
\end{eqnarray}
\end{prop}

\proof
A representation $(\theta,s)$ of $D(SU(2))$ is simple with respect to the action of the group $SU(2)$ if and only if ${\cal V}_{\theta,s}$ admits a non zero $SU(2)$ invariant vector $\omega$. Such a vector $\omega$ satisfies the condition:
\begin{eqnarray}
\forall \; x\in SU(2)\;,\;\; [\rho^\theta_s(x\otimes 1)\omega] \; = \; \omega\;.
\end{eqnarray}
From the expression (\ref{representation of D(SU(2))}) of the representation, it is clear that such an invariant vector $\omega$ exists if and only if $s=0$. In that case, the invariant vector is the unit element $\omega=1$.

Given a simple representation labelled by $\theta$, the symmetric propagator  is defined as:
\begin{eqnarray}
K_\theta \; = \; \int dg \; \overline{\omega(g)}\; [\rho^\theta(\cdot) \; \omega](g)\; \in D(SU(2))^*.
\end{eqnarray}
Using the expression of the representation (\ref{representation of D(SU(2))}), one sees immediately that the kernel is given by the above formula (\ref{Kernel of SU(2)}). The expansion in Fourier modes of the kernel is a direct consequence of the following expansion of the characteristic function  $\delta_\theta$:
\begin{eqnarray}
\delta_\theta(g) \; = \; \sum_{j\in {\mathbb N}}\chi_j(h_\theta) \; \chi_j(g) \;.
\end{eqnarray}
As a consequence, the symmetric propagator  is the Fourier transform of $SU(2)$ characters evaluated on irreducible unitary finite dimensional representations. The propagator is obtained immediately from the expression of the antipode on $D(SU(2))^*$. $\Box$

\medskip

Remark that the compact group case works formally exactly as the finite group case at the technical level as well as the conceptual level. Duality relations are also very similar to the finite group case  and are presented in the following theorem.

\begin{theo}[Duality in the case of $D(SU(2))$]
Let $T$ be a tetrahedron whose edges $(IJ)$ are colored by simple representations $(\theta_{IJ})$
 of $D(SU(2)).$ The evaluation of this spin network is given by: 
\begin{eqnarray}
T(\theta_{IJ}) \; = \; \sum_{(j_{IJ})\in{\mathbb N}^{\times 6}}\prod_{IJ} \chi_{j_{IJ}}(\theta_{IJ}) \left\{\begin{array}{ccc}
j_{{12}} & j_{{13}} & j_{{14}}\\
 j_{{34}}& j_{{24}} & j_{{23}}
\end{array}\right\}^2\;\text{with}\;\; \chi_j(\theta)=\frac{\sin(j+1)\theta}{\sin\theta}\label{evaluationcompactgroup}\;.
\end{eqnarray}
Moreover, $T(\theta_{IJ})$ can be expressed in term of $(6j)$ symbols between simple representations of $DSU(2)$ as follows
\be
T(\theta_{IJ}) \; = \;\frac{\pi^3}{32}\left\{\begin{array}{ccc}
\theta_{{12}} & \theta_{{13}} & \theta_{{14}}\\
 \theta_{{34}}& \theta_{{24}} & \theta_{{23}}
\end{array}\right\},
\ee
where the  $(6j)$ symbol  is 
normalized with respect to the Plancherel measure $ d\theta \sin^2 \theta$ as shown in the  appendix.
This $(6j)$ symbol satisfies a duality relation with the square of the normalised $(6j)$ symbol of
$\SU(2)$:
\begin{eqnarray}\label{duality for compact group}
\left\{\begin{array}{ccc}
\theta_{{12}} & \theta_{{13}} & \theta_{{14}}\\
 \theta_{{34}}& \theta_{{24}} & \theta_{{23}}
\end{array}\right\}
\; = \; \frac{32}{\pi^3}\sum_{j_{IJ}} \left( \prod_{I<J}
( \sin((j_{IJ} +1)\theta_{IJ})\right)
\left\{\begin{array}{ccc}
j_{{12}} & j_{{13}} & j_{{14}}\\
 j_{{34}}& j_{{24}} & j_{{23}}
\end{array}\right\}^2.
\end{eqnarray}

Furthermore, $(6j)$ symbols of simple representations of $D(SU(2))$ can be computed  and are given by:
\begin{eqnarray}\label{6j of DSU(2)}
\left\{\begin{array}{ccc}
\theta_{{12}} & \theta_{{13}} & \theta_{{14}}\\
 \theta_{{34}}& \theta_{{24}} & \theta_{{23}}
\end{array}\right\}
\; = \; \frac{1}{\pi} \frac{1}{\sqrt{\det[\cos(\theta_{\widehat{IJ}})]}}\;.
\end{eqnarray}
As a result, the series (\ref{duality for compact group}) converges when the determinant of the matrix $(\cos(\theta_{\widehat{IJ}})_{IJ})$ does not vanish.
\end{theo}

\proof
To prove this theorem, we use the same techniques as in the finite group case. First, we evaluate the simple spin-network associated to the tetrahedron, then we relate this evaluation to $(6j)$ symbols computing the normalisation of the symmetric intertwiner, finally we compute explicitly the expression of the $(6j)$ symbols from relation (\ref{dualityclass2}).

The two first step mimic the finite group case. Indeed, the evaluation $T(\theta_{IJ})$ of the simple spin-network associated to the tetrahedron graph is obtained by integrating over $D(SU(2))^{\otimes 4}$ the function (\ref{Tetrahedron function}) where representations $(A_{IJ})$ are replaced by $(\theta_{IJ})$ and the kernel $K_\theta$ is the one given in the previous proposition (\ref{Proposition compact group}). Therefore, a direct calculation shows the formula (\ref{evaluationcompactgroup}).

As a result the normalized $(6j)$ symbol satisfies the following equality:
\begin{eqnarray}
\frac{\pi}{2}\left\{\begin{array}{ccc}
\theta_{{12}} & \theta_{{13}} & \theta_{{14}}\\
 \theta_{{34}}& \theta_{{24}} & \theta_{{23}}
\end{array}\right\}
\; = \;
\frac{T(\theta_{IJ})}{\sqrt{\Theta(\theta_{14}, \theta_{13}, \theta_{12})}\sqrt{\Theta(\theta_{13}, \theta_{24}, \theta_{23})}\sqrt{\Theta(\theta_{13}, \theta_{23}, \theta_{24})}\sqrt{\Theta(\theta_{14}, \theta_{34}, \theta_{24})}}\nonumber
\end{eqnarray}
where the explicit expression for $\Theta$ is given in Appendix A and reads
\begin{eqnarray}
\Theta(\alpha,\beta,\gamma) \; = \; \frac{\pi}{4 \sin \alpha \sin \beta \sin \gamma } Y({\alpha},{\beta},{\gamma}).
\end{eqnarray}

Finally, one ends up with the duality relation (\ref{duality for compact group}) between simple $(6j)$ symbols of $D(SU(2))$ and $(6j)$ symbols of unitary representations of $SU(2)$.
We finish the proof by using the expression (\ref{dualityclass2}).
 $\Box$

\medskip

In \cite{KM} the Clebsch-Gordan maps of unitary representations of $D(SU(2))$ have been explicitely computed. We could use their expression to give an alternative derivation of the expression (\ref{6j of DSU(2)}).

\subsubsection{The quantum group case $U_q(so(3,1))$}

In this section we will find the analog of Barrett relation (\ref{dualityqroot}) in the case where $q$ is real. As in the previous cases, we will show that this amounts to express the evaluation of  a relativistic tetrahedral spin network of $U_q(sl(2,{\mathbb C}))$  colored by simple  principal representations in terms of Fourier components of   $U_q(su(2))$. 
We will use the definition of $U_q(sl(2,{\mathbb C}))$ (that we will sometimes abusively call the quantum Lorentz group denoted   $U_q(so(3,1))$), as being the quantum double of $U_q(su(2)))$. 
We will use the notations of \cite{BR1}\cite{BR2}\cite{NR}.

Let $q=e^{-\kappa}\in ]0,1[$, and consider the quantum $\star$-Hopf algebra $U_q(so(3,1))$ defined as in \cite{BR1} to be the quantum double of $U_q(su(2))$. If $z\in {\mathbb C}$ we denote $[z]_q=\frac{q^z-q^{-z}}{q-q^{-1}}.$
We start by recalling basic facts about representation theory of $U_q(so(3,1))$. Principal unitary irreducible representations of $U_q(so(3,1))$ have been classified and
are labelled by a couple $\alpha=(k,i\rho)$ with $k\in {\mathbb Z}$ and
$\rho\in {\mathbb R}.$  Note that the representation $(k,i\rho)$ is unitary equivalent to $(-k,-i\rho)$ and to $(k,i(\rho+\frac{2\pi}{\kappa})).$
 $U_q(so(3,1))$ being the quantum double of $U_q(su(2))$ it contains it as a sub-Hopf-algebra. As usual, the simple  principal representations are those which admit a non-zero $U_q(su(2))$ invariant vector; this is equivalent to the condition $k=0$ \cite{NR}. We will denote by $\omega$ this invariant unit vector. 
Recall that the algebra of functions on the quantum hyperboloid is defined as a pile of fuzzy sphere, i.e.
 $Fun({\cal H}_{3q})=\oplus_{j\in {\mathbb N}} Mat_{j+1}(\mathbb C)$ where $Mat_{j+1}(\mathbb C)$ is the algebra of matrices of size $j+1$ with unit element $\mathbb I_{j+1}$.
The Haar measure on $ F_q(SO(3,1))$ 
descends to $Fun({\cal H}_{3q})$  and if 
 $f$ is an element of $Fun({\cal H}_{3q})$, $f=\oplus_j f_j,$ the Haar measure 
$h(f)=\sum_j [j+1]_q tr(f_j q^{-2J_3})$ where $q^{2J_3}$ is the Cartan generator of $U_q(su(2)).$

\begin{prop}[Symmetric propagator of the quantum Lorentz group]
Given a simple representation $\rho$, the symmetric propagator   $K_\rho$ is a function on the quantum hyperboloid ${\cal H}_{3q}$, i.e. $K_\rho \in Fun({\cal H}_{3q}) \subset F_q(SO(3,1))$. Right and left $U_q(su(2))$ coinvariance of $K_\rho$ implies that $K_\rho$ is diagonal in the sense that:
\begin{equation} \label{quantumkernel}
K_\rho=\sum_{j\in {\mathbb N}} K_\rho(j)\mathbb I_{j+1} \;\;\;\; \text{with}\;\; K_\rho(j)=\frac{[i(j+1)\rho]_q}{[i\rho]_q[j+1]_q}\;.
\end{equation}
The  propagator is defined as $K_\rho(x,y) \equiv (id\otimes S)\Delta K_\rho$, is a two-point function on the quantum hyperboloid and is expressed explicitely in terms of  $U_q(su(2))$ Clebsch-Gordan coefficients  through the formula (97) of \cite{NR}. 
\end{prop}

The results presented in this proposition have already been proven in \cite{BR2}\cite{NR}. It is interesting to remark that $K_\rho$ can be viewed as the Fourier transform of some function on the quantum hyperboloid when written as follows:
\begin{eqnarray}\label{Quantum kernel}
K_\rho \; = \; \frac{1}{\sin \kappa \rho}\sum_{j=0}^{+\infty} \sin(\kappa \rho (j+1)) \frac{\mathbb I_{j+1}}{[j+1]_q}\;.
\end{eqnarray}
There is a strong resemblance with the expression (\ref{Kernel of SU(2)}) in the compact group case: up to a constant $K_\rho$ is the Fourier transform of the function $\mathbb I_{j+1}/[j+1]_q$. The fact that $K_\rho$ can be expressed as a Fourier transform is the key property to establish duality relations.

However, proving duality relations in that case is a bit more involved than the other cases because of the non-compactness of the quantum group and the fact that $(6j)$ symbols of simple representations of the quantum Lorentz group are still unknown. The duality relation we are going to establish can be viewed as giving an explicit expression  for  $(6j)$ symbols of simple principal representations  of $U_q(so(3,1))$.

In the following, we first compute the evaluation of the tetrahedron simple spin-network, then we establish the relation between this evaluation and simple $(6j)$ symbols of the quantum Lorentz group.

\begin{prop}[Duality in the case of the quantum Lorentz group $U_q(so(3,1))$]\label{quantumevaluationlorenztian}
Let $T$ be a tetrahedron whose edges $(IJ)$ are colored with simple representations $\rho_{IJ}$. The evaluation of the simple spin-network associated to $T$ is a function of the six variables $\rho_{IJ}$ that is given by the following formula:
\begin{eqnarray}
T(\rho_{IJ}) \; = \; \sum_{(j_{IJ})\in {\mathbb N}^{\times 6}}
\prod_{I<J}\frac{\sin \left(\kappa (j_{IJ}+1)\rho_{IJ}\right)}{\sin \left(\kappa \rho_{IJ}\right)}\left\{\begin{array}{ccc}
j_{\widehat{12}} & j_{\widehat{13}} & j_{\widehat{14}}\\ j_{\widehat{34}}& j_{\widehat{24}} & j_{\widehat{23}}
\end{array}\right\}^2_q.\label{evallorentzienquantique}
\end{eqnarray}
 This series is uniformly and absolutely convergent in $\rho_{IJ}.$
As a result it is a continuous function of $\rho_{IJ}$ which is $\frac{2\pi}{\kappa}$ periodic in each variable.
\end{prop}

\proof
In \cite{NR} we have given a procedure for evaluating this simple spin network. In particular, we used  a regularisation to deal with the non-compactness of the quantum group.

To evaluate $T(\rho_{IJ})$, we first construct an element $F_T(\rho_{IJ}) \in Fun({\cal H}_{3q})^{\otimes 4}$ exactly in the same way as in the previous cases. In particular, $F_T(\rho_{IJ})$ takes the form (\ref{Tetrahedron function}) where representations are replaced by simple representations of $U_q(so(3,1))$ and the kernel is given by (\ref{Quantum kernel}). Note that $F_T(\rho_{IJ})$ is now a non-commutative Feynman graph.

\begin{figure}
\psfrag{m1}{$j_{12}$}
\psfrag{m2}{$j_{23}$}
\psfrag{m3}{$j_{34}$}
\psfrag{m4}{$j_{14}$}
\psfrag{m5}{$j_{24}$}
\psfrag{m6}{$j_{13}$}
\psfrag{=}{$=A(j_{IJ})$}
\centering
\includegraphics[scale=0.5]{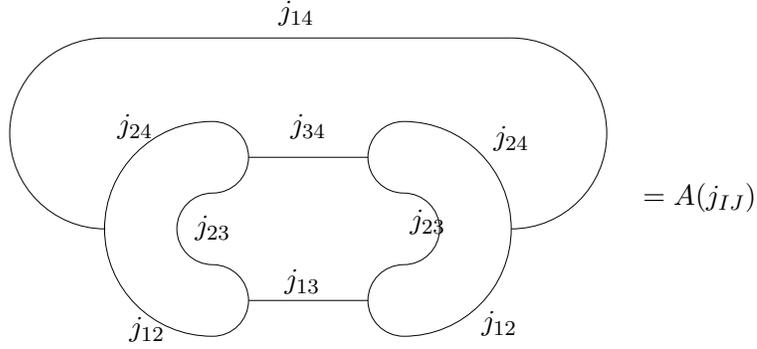}
\caption{Definition of $A(j_{IJ})$}
\label{grapheintermediaire}
\end{figure}

The evaluation of this simple spin network is obtained through a regularisation procedure which consists in choosing one point out of the four vertices and then integrating over only the remaining 3 vertices. This regularisation does not depend on the choice of the chosen vertex. For simplicity purpose in the computations, we choose not to integrate over the third point and, as a result, we have:
\begin{eqnarray}
T(\rho_{IJ}) 1 =(h \otimes id\otimes h\otimes h) \; F_T(\rho_{IJ})\;.
\end{eqnarray}
In that expression, $h$ denotes the Haar measure on the quantum hyperboloid \cite{NR}.
The obtained function is a constant whose value $T(\rho_{IJ})$ is obtained by applying the counit to it:
\begin{eqnarray}
T(\rho_{IJ}) & = & h(K_{\rho_{14}(1)}K_{\rho_{12}(1)}K_{\rho_{13}(1)}) \; h(S(K_{\rho_{13}(2)}) S(K_{\rho_{23}(2)}) K_{\rho_{34}(1)}) \nonumber \\
&&h(S(K_{\rho_{34}(2)} K_{\rho_{24}(2)} K_{\rho_{14}(2)})).
\end{eqnarray}
We can perform the integrations using the same technique as explained in 
\cite{NR} and we obtain the following result:
\begin{equation}
T(\rho_{IJ})=\sum_{j_{IJ}}\prod_{I<J}K_{\rho_{IJ}}(j_{IJ})
[j_{12}+1]_q[j_{23}+1]_q[j_{24}+1]_q(-1)^{(j_{12}+j_{23}+j_{24})} A(j_{IJ})
\end{equation}
where $A(j_{IJ})$ is the evaluation of the $U_q(su(2))$ spin-network whose colored graph is depicted in figure (\ref{grapheintermediaire}).
Using usual rules, we can evaluate $A(j_{IJ})$ and we obtain:
\begin{equation}
A(j_{IJ})=(-1)^{(j_{12}+j_{23}+j_{24})}
\frac{\prod_{I<J}[j_{IJ}+1]_q}{[j_{12}+1]_q[j_{23}+1]_q[j_{24}+1]_q}
\sixjsym{j_{34}}{j_{24}}{j_{23}}{j_{12}}{j_{13}}{j_{14}}^2_q.
\end{equation}
As a result we obtain the announced proposition on the expression of
$T(\rho_{IJ})$.

It remains to show that this series is uniformly convergent.
We first use the fact that, due to orthogonality relations satisfied by the
$(6j)$ we have:
\begin{eqnarray}
\sixjsym{j_{13}}{j_{14}}{j_{12}}{j_{34}}{j_{23}}{j_{24}}^2_q
\leq \text{min}(\frac{1}{[j_{12}+1]_q[j_{24}+1]_q},\frac{1}{[j_{13}+1]_q[j_{34}+1]_q},\frac{1}{[j_{14}+1]_q[j_{23}+1]_q}).
\end{eqnarray}
Define $n=\sum_{(IJ)}j_{IJ}\in {\mathbb N},$ necessarily there exists a couple $(IJ)$ such that $j_{IJ}\geq n/6.$ As a result we obtain that the previous bound implies the following:
\begin{equation}
\sixjsym{j_{13}}{j_{14}}{j_{12}}{j_{34}}{j_{23}}{j_{34}}^2_q\leq
[j_{IJ}+1]_q^{-1}\leq [\frac{1}{6}n+1]_q^{-1}
\end{equation}
To prove the uniform convergence, it is sufficient to show absolute convergence of the series. And we have:
\begin{eqnarray}
&&\sum_{(j_{IJ})}\left\{\begin{array}{ccc}
j_{34} & j_{24} & j_{23}\\ j_{12}& j_{13} & j_{14}
\end{array}\right\}^2_q
\prod_{I<J}\vert\frac{\sin \left(\kappa (j_{IJ}+1)\rho_{IJ}\right)}{\sin \left(\kappa \rho_{IJ}\right)}\vert\\
&&\leq \sum_{(j_{IJ})}\left\{\begin{array}{ccc}
j_{34} & j_{24} & j_{23}\\ j_{12}& j_{13} & j_{14}
\end{array}\right\}^2_q\prod_{I<J} (j_{IJ}+1)\\
&&\leq \sum_{n\geq 0}{[n/6+1]_q}^{-1}A(n)(n+1)^6.
\end{eqnarray}
In the last equality, we have introduced $A(n)$ as the number of sixtuplet  $(j_{IJ})$ satisfying
$n=\sum_{(IJ)}j_{IJ}.$ We trivially have $A(n)\leq n^6.$
Therefore because of the exponential growth of $[n/6+1]_q$ we obtain that the series is absolutely convergent. As a result it is a continuous function of the argument $\rho_{IJ}.$ $\Box$

\medskip

Remark: The continuity property of this function is in sharp contrast with the classical case. The limit $\kappa$ goes to zero with $\kappa\rho_{IJ}$ fixed to $\theta_{IJ}$ in the series (\ref{evallorentzienquantique}), can be formally evaluated by inverting the limit and the summation (\ref{6j of DSU(2)}), and is given by:
\begin{equation}
\frac{\pi^2}{32}
\prod_{I<J}\frac{1}{\sin(\t_{IJ})}
 \frac{1}{\sqrt{\det[\cos\t_{IJ}]}}
\end{equation}
which exhibits numerous divergences.

\medskip

Now, the point is to show that $T(\rho_{IJ})$ can be expressed as $(6j)$ symbols of principal simple representations of the quantum Lorentz group $U_q(so(3,1))).$ To do so, we first need to introduce a precise notion of $(6j)$ symbols of $U_q(so(3,1)).$ The Clebsch-Gordan maps and the Plancherel measure have already been studied in \cite{BR1,BR2}.
Using the appendix  of the present work we obtain:

\begin{theo}[Duality in the case of a quantum group]
We consider the $(6j)$ coefficients of simple principal representations of the
quantum Lorentz group  normalised with respect to the Plancherel
measure $ d\rho\, [i\rho]_q^{2}$.
These coefficients satisfy the following identity:
\begin{eqnarray}\label{dualityUqSL}
\left(\frac{\pi}{2\kappa}\right)^{3}
\prod_{K=1}^{4}{\sqrt{ \Theta_K(\rho_{IJ})}}
\left\{\begin{array}{ccc}
 \rho_{{12}}& \rho_{{13}} & \rho_{{14}}\\
\rho_{{34}} & \rho_{{24}} & \rho_{{23}}
\end{array}\right\}
\; = \sum_{(j_{IJ})\in {\mathbb N}^{\times 6}}
\left( \prod_{I<J} \sin (\kappa (j_{IJ}+1)\rho_{IJ}) \right)
\left\{\begin{array}{ccc}
j_{\widehat{12}} & j_{\widehat{13}} & j_{\widehat{14}}\\
 j_{\widehat{34}}& j_{\widehat{24}} & j_{\widehat{23}}
\end{array}\right\}_q^2
\end{eqnarray}
where 

\begin{eqnarray}
\Theta(\rho_1,\rho_2,\rho_3)&=&\sum_{n=1}^{+\infty}\frac{1}{[n]_q}
\prod_{k=1}^3\frac{\sin(n\kappa\rho_{k})}{\sin(\rho_{k}\kappa)}\\
&=&  \frac{1}{8} \frac{\vartheta_1(\kappa\rho_1)\vartheta_1(\kappa\rho_2)\vartheta_1(\kappa\rho_3)}{\sin(\kappa\rho_1) \sin(\kappa\rho_2) \sin(\kappa\rho_3)}\frac{q^{1/4}(q^{-1}-q)(q^2;q^2)^3_{\infty}}{\vartheta_4(\kappa\frac{\rho_1+\rho_2+\rho_3}{2})\vartheta_4(\kappa\frac{-\rho_1+\rho_2+\rho_3}{2})\vartheta_4(\kappa\frac{\rho_1-\rho_2+\rho_3}{2})\vartheta_4(\kappa\frac{\rho_1+\rho_2-\rho_3}{2})},\nonumber
\end{eqnarray}
with  $\vartheta_1$ and $\vartheta_4$ are Jacobi functions whose definitions are recalled in the appendix.

\end{theo}

\proof
It is a direct consequence of the proposition of the appendix and the previous 
proposition.$\Box$

\medskip

We have shown how duality relations between $(6j)$ symbols in different exemples. The following sections aims at presenting these relations in a
more general context.



\subsection{Examples of duality between  simple $(6j)$ of the double cross product and simple $(6j)$ of the associated  bicrossproduct}
In this section we give a brief review of the notion of double cross product and bicross product of two Hopf algebras. We then show that to each of the Hopf algebras $D$  studied in the previous section there is an associated Hopf algebra $\tilde{D}$ which structure will be analysed. We describe the simple representations of  $\tilde{D}$ and express the duality relation between $(6j)$ of simple representations of $D$ with $(6j)$ of simple representations of $\tilde{D}.$

We first recall the notion of double cross product, which generalizes the notion of quantum double.

Let $B,C$ two Hopf algebras endowed with  actions $\alpha,\beta:$
\begin{eqnarray}
&&\alpha:C\otimes B\rightarrow C,\;\; c\otimes b\mapsto c\triangleleft b\\
 &&\beta: C\otimes B\rightarrow B,\;\; c\otimes b\mapsto c\triangleright b
\end{eqnarray}
such that  the compatibility relations of Definition 7.2.1 of \cite{Ma} are satisfied.

The double cross product Hopf algebra is the Hopf algebra denoted $B\bowtie C$ built on the vector space $B\otimes C$ with product and coproduct defined as:
\begin{eqnarray}
&&(b\otimes c)(b'\otimes c')=\sum_{(c)(b')}b(c_{(1)}\triangleright b'{}_{(1)})
\otimes (c_{(2)}\triangleleft b'{}_{(2)})c'\\
&&\Delta(b\otimes c)= \sum_{(b)(c)}b_{(1)}\otimes c_{(1)}\otimes b_{(2)}\otimes c_{(2)}.
\end{eqnarray}

We have already encountered two particular case of this notion.

In the classical case  let  $H$ be a Poisson Lie group, define $B={\mathbb C}[H]$, and
 $C={\mathbb C}[H^*].$ We have $D=\mathbb{C}[{\cal D}]=B\bowtie C$ with the action $\alpha$ and $\beta$ being the dressing actions.

In the quantum case , let $A$ be a Hopf algebra, the quantum double is the Hopf algebra  $D=D(A)=A\bowtie A^{\star cop}$ where the action $\alpha$ and $\beta$ are the mutual coadjoint actions.

We now recall the notion of bicrossproduct of two Hopf algebras.

Let $B,C$ two Hopf algebras endowed with  action $\gamma$ and coaction $\delta$
\begin{eqnarray}
&&\gamma:C\otimes B\rightarrow B,\;\; c\otimes b\mapsto c\triangleright b\\
 &&\delta: C\rightarrow C\otimes B,\;\; c\mapsto \sum_{(c)}c^{(\bar{1})}\otimes c^{(\bar{2})}
\end{eqnarray}
such that  the compatibility relations of Theorem 6.2.2 of \cite{Ma} are satisfied.

The bicross product Hopf algebra is the Hopf algebra denoted $B\lrbicross C$ built on the vector space $B\otimes C$ with product and coproduct defined as:
\begin{eqnarray}
&&(b\otimes c)(b'\otimes c')=
\sum_{(c)}b(c_{(1)}\triangleright b')
\otimes c_{(2)}c'\\
&&\Delta(b\otimes c)= \sum_{(b)(c)}b_{(1)}\otimes c_{(1)}^{(\bar{1})}\otimes b_{(2)}c_{(1)}^{(\bar{2})}\otimes c_{(2)}.
\end{eqnarray}

An important theorem is the proposition 7.2.4 of \cite{Ma} which states that the compatibility relations for $\alpha,\beta$  necessary for the construction of $B\bowtie C$ are equivalent to the compatibility relations necessary for the construction of $C^*\lrbicross B$ where the corresponding action and coaction are  obtained from $\alpha$ and $\beta$ by dualisation of $C.$

To each of the cases studied in the previous section we associate the Hopf algebra 
$D=B\bowtie C.$ We can therefore define the Hopf algebra $\tilde{D}=C^*\lrbicross B.$
We will now study the simple representations of $\tilde{D}$ and compute the associated $(6j).$

In the classical case, if $H$ is a Poisson Lie group, then the groups $H$ and $H^*$ form a right-left matched pair in the sense of \cite{Ma} (Definition 6.2.10) and we denote $\alpha$ and $\beta$ the dressing actions.
We define $D={\mathbb C}[{\cal D}(H)]={\mathbb C}[H]\bowtie {\mathbb C}[H^*]$ which is a cocomutative Hopf algebra.

We can define $\tilde{D}=F(H^*)\lrbicross {\mathbb C}[H]$. This is a cocommutative Hopf algebra if and only if $H^*$ is abelian i.e if $H$ is endowed with its nul Poisson structure.
 The structure of $F(H^*)\lrbicross {\mathbb C}[H]$ is explicitely  described in Example 6.2.11 of \cite{Ma} and reads in the  basis $\delta_{\gamma}\otimes g$
\begin{eqnarray}
&&(\delta_{\gamma}\otimes g)(\delta_{\gamma'}\otimes g')=
\delta_{\alpha_{g}(\gamma),\gamma'}(\delta_{\gamma}\otimes g g'),\\
&&\Delta(\delta_{\gamma}\otimes g)=
\sum_{\gamma\gamma'=\gamma''}\delta_{\gamma}\otimes \beta_{\gamma'}(g)\otimes
 \delta_{\gamma'}\otimes g.
\end{eqnarray}

We will now study the case where $H=SU(2).$ 
\subsubsection{The Euclidean group and the Lorentz group}
We take $H=SU(2)$ with its nul Poisson structure. We have 
$D=\mathbb{C}[SU(2)\ltimes {\mathbb R}^3]$ and 
 $\tilde{D}=F(\mathbb{R}^3)\lrbicross {\mathbb C}[SU(2)].$
Because $\tilde{D}$ is cocommutative the following map is an isomorphism of Hopf algebras:
\begin{eqnarray}
&&\mathbb{C}[{\mathbb R}^3\rtimes SU(2)]\rightarrow F(\mathbb{R}^3)\lrbicross {\mathbb C}[SU(2)]\\
&& (\vec{p},u)\mapsto \int  e^{i\vec{x}.\vec{p}}\;\;\delta_{\vec{x}}\otimes u
 d^3 \vec{x},
\end{eqnarray} 
where ${\mathbb R}^3\rtimes SU(2)$ is endowed with the group law 
$(\vec{p},u)(\vec{p'},u')=(\vec{p}+\Lambda(u)\vec{p'},uv).$

We now take $H=SU(2)$ with its non zero standard Lie Poisson structure, we have $AN=H^{*}$ and $D=\mathbb{C}[SL(2,\mathbb{C})]=\mathbb{C}[SU(2)]\bowtie
 \mathbb{C}[AN].$
 As a result we have $\tilde{D}=F(AN)\lrbicross \mathbb{C}[SU(2)].$
(****)
\subsubsection{The case of a quantum double}
Let $A$ be a Hopf algebra and define  $D=D(A)=A\bowtie A^{*cop}$ to be the quantum  double of $A.$ 
We have $\tilde{D}= A_{op}\lrbicross A$ and it has been shown by S.Majid (Theorem 6.2.9 of \cite{Ma}) that
 $\tilde{D}$ is isomorphic to $A^{op}\otimes A$ as a Hopf algebra.  Under this isomorphism the subalgebra $1\lrbicross A$ is sent to the diagonal embbedding of $A$ in $A^{op}\otimes A$ i.e the subalgebra $(S\otimes id)\Delta(A).$ 
As a result the finite dimensional representations of $\tilde{D}$ which are simple consists in the representations of $A^{op}\otimes A$ of the form ${}^{t}\pi\otimes \pi$ where $\pi$ is an irreducible finite dimensional representation of $A.$ 
As a result when no multiplicities are involved i.e $A={\mathbb C}[SU(2)]$ or 
$A=U_q(su(2))),$
the $(6j)$ of simple representations ${}^t\pi_{IJ}\otimes \pi_{IJ}$ of $\tilde{D}$ are the square of the $(6j)$ of the representations $\pi_{IJ}.$
The duality relations that one obtains in the case $A={\mathbb C}[SU(2)]$ or 
$A=U_q(su(2)))$ can therefore be understood as a duality between $(6j)$ of unitary irreducible representations (infinite dimensional) of $D$ and $(6j)$ of unitary irreducible representations (finite dimensional) of  $\tilde{D}.$

 \section{Conclusion}
In this work we have analyzed the duality relation between $6j$ symbols of simple representations for different exemples which are all associated to the double cross product construction. We have shown that the duality relation expresses the relation between the $6j$ of simple representations of $D$ with the $6j$ of simple representations of $\tilde{D}.$ Although at this stage it just amounts to check this on some examples we firmly believe that these results can  be extended to a larger framework where $D$ 
encompasses more cases than the one considered in our work. The largest class one could think of is $D=H_1\bowtie H_2$ where $H_1,H_2$ are Hopf algebras and 
$\tilde{D}= H_2^*\lrbicross H_1.$
The duality relation would therefore expresses a relation  between  the evaluation of  the tetrahedral  spin network  of  $D$ with the evaluation  of tetrahedral spin network of $\tilde{D}.$ In order to obtain such relation one has first to classify simple representations of $D$ and $\tilde{D},$ only partial results being available in the litterature.

The duality relations can be also studied from a geometrical point of view because the large spin expansion of the $6j$ is related to the geometry of an associated tetrahedron. One can show that all the duality relation which have been obtained in our work have a geometrical meaning.

\appendix
\section{Conventions and results on Clebsch-Gordan and 6j coefficients}
In this appendix we give the convention of normalization that we use in our work.
We will only deal with the cases $ ISU(2), SL(2,\mathbb C), D(SU(2)), U_q(so(3,1))$. In these cases the space of interwiners from one irrep to the tensor product of two irreps is of dimension $0$ or $1$ and therefore there is no need to introduce multiplicities in the definition of Clebsch-Gordan maps and $(6j)$ symbols. However  Clebsch-Gordan maps have to be properly normalized and we will concentrate now on this point.
We use the following normalization of Plancherel measure for simple representations:
\begin{enumerate}
\item  $ISU(2), P(m)=m^2, m\in {\mathbb R}^+$
\item $SL(2,\mathbb C), P(\rho)=\rho^2, \rho\in  {\mathbb R}^+$
\item $D(SU(2)), P(\theta)=\sin^2 \theta, \theta\in [0,\pi]$
\item $U_q(so(3,1)),  P(\rho)=[i\rho]_q^2, \rho\in [0,\frac{\pi}{\kappa}[,
 q=e^{-\kappa}, \kappa>0.$
\end{enumerate}
Let $\rho_1,\rho_2,\rho_3$ be simple irreducible unitary representations  of one of the above Hopf algebras,
because $P$ is up to a scalar the Plancherel measure,
we obtain:
\begin{equation}
\iota^s[\rho_1;\rho_2]=N\; id_{V_{\rho_1}}\frac{\delta(\rho_1-\rho_2)}{P(\rho_1)},
\end{equation}
where $N$ is a normalization constant independent of $\rho_i$
and $V_{\rho_1}$ is the vector space of the representation $\rho_1.$
We can  define a function $\Theta(\rho_1,\rho_2,\rho_3)$  as:
\begin{equation}
\Theta(\rho_1,\rho_2,\rho_3)=\langle\omega\vert\iota^s[\rho_1\otimes \rho_2;\rho_3]\vert
 \omega^{\otimes 2}\rangle.
\end{equation}

We can therefore define the normalized intertwiner as:
\begin{equation}
\psi_{\rho_1,\rho_2}^{\rho_3}=\frac{1}{\sqrt{N \Theta(\rho_1,\rho_2,\rho_3)}}
\iota^s[\rho_1\otimes\rho_2;\rho_3],
\end{equation}
it satisfies:
\begin{equation}
\psi_{\rho_1,\rho_2}^{\rho_3} (\psi_{\rho_1,\rho_2}^{\rho_3'})^{\dagger}=
\frac{\delta(\rho_3-\rho_3')}{P(\rho_3)},
\end{equation}
as well as 
\begin{equation}
\langle\omega\vert\psi_{\rho_1,\rho_2}^{\rho_3}\vert
 \omega^{\otimes 2}\rangle=(\frac{ \Theta(\rho_1,\rho_2,\rho_3)}{N})^{1/2}.
\end{equation}

The $(6j)$ symbol is constructed through the use of normalized interwiner, namely it is defined by the equation:
\begin{equation}
\psi_{\rho_{34},\rho_{14}}^{\rho_{24}}
\psi_{\rho_{13},\rho_{12}}^{\rho_{14}}
(\psi_{\rho_{34},\rho_{13}}^{\rho_{23}})^{\dagger}
=\sixjsym{\rho_{12}}{\rho_{13}}{\rho_{14}}{\rho_{34}}{\rho_{24}}{\rho_{23}}
\psi_{\rho_{23},\rho_{12}}^{\rho_{24}}.
\end{equation}

\begin{prop}

Let $T$ be the tetrahedron graph colored with representations $\rho_{IJ},$ we have the following result:
\begin{equation}
I(T)=N\prod_{K=1}^4 \sqrt{\Theta_K(\rho_{IJ})})
 \sixjsym{\rho_{12}}{\rho_{13}}{\rho_{14}}{\rho_{34}}{\rho_{24}}{\rho_{23}},
\end{equation}
where we have denoted $\Theta_1(\rho_{IJ})=\Theta(\rho_{12},\rho_{13},\rho_{14})$ and a similar definition for $\Theta_K.$
\end{prop}
\proof
\begin{eqnarray}
I(T)&=&\langle\omega \vert\iota^s[\rho_{34}\otimes\rho_{14};\rho_{24}]
\iota^s[\rho_{13}\otimes\rho_{12},\rho_{14}]
(\iota^s[\rho_{34}\otimes\rho_{13},\rho_{23}])^{\dagger} \vert\omega^{\otimes 2}\rangle\\
&=& (N^3\Theta_4\Theta_1\Theta_3)^{1/2}
\langle\omega\vert\psi_{\rho_{34},\rho_{14}}^{\rho_{24}}
\psi_{\rho_{13},\rho_{12}}^{\rho_{14}}
(\psi_{\rho_{34},\rho_{13}}^{\rho_{23}})^{\dagger}\vert\omega^{\otimes 2}\rangle\\
&=& (N^3\Theta_4\Theta_1\Theta_3)^{1/2}\sixjsym{\rho_{12}}{\rho_{13}}{\rho_{14}}{\rho_{34}}{\rho_{24}}{\rho_{23}}\langle\omega\vert\psi_{\rho_{23},\rho_{12}}^{\rho_{24}}\vert \omega^{\otimes 2}\rangle\\
&=&N\prod_{K=1}^4 \sqrt{\Theta_K(\rho_{IJ})})
 \sixjsym{\rho_{12}}{\rho_{13}}{\rho_{14}}{\rho_{34}}{\rho_{24}}{\rho_{23}}.
\end{eqnarray}
$\Box.$

We now give the explicit expressions for $N$ and $\Theta$ in the 4 considered cases.

\noindent

1. $ISU(2),\;N=2\pi^2$
\begin{equation}
\Theta(m_1,m_2,m_3)=\frac{4\pi}{m_1m_2m_3}\int_0^{+\infty}
 \sin(m_1 a)\sin(m_2 a)\sin(m_3 a)\frac{da}{a}=
\frac{\pi^2}{m_1 m_2 m_3} \; Y(m_1,m_2,m_3),
\end{equation}
where $Y$ is the characteristic function of the set $\Delta$ defined as
\begin{equation}
\Delta=\{(a,b,c)\in ({\mathbb R}^+)^3,a\leq b+c, b\leq a+c, c\leq a+b\}.
\end{equation}
2. $SL(2,\mathbb C),\;N=2\pi^2$
\begin{eqnarray}
&&\Theta(\rho_1,\rho_2,\rho_3)=
\frac{4\pi}{\rho_1\rho_2\rho_3}\int_0^{+\infty}
 \sin(\rho_1 r)\sin(\rho_2 r)\sin(\rho_3 r)\frac{dr}{\sinh r}=\\
&&\frac{\pi^2}{2\rho_1\rho_2\rho_3}(
\tanh(\frac{\pi}{2}(\rho_1+\rho_2-\rho_3))+\tanh(\frac{\pi}{2}(\rho_1-\rho_2+\rho_3))+\\
&&\tanh(\frac{\pi}{2}(-\rho_1+\rho_2+\rho_3))-
\tanh(\frac{\pi}{2}(\rho_1+\rho_2+\rho_3))).
\end{eqnarray}
3. $D(SU(2)),\;N=\frac{\pi}{2} $
\begin{equation}
\Theta(\theta_1,\theta_2,\theta_3)=\frac{\pi}{4}\frac{Y(\theta_1,\theta_2,\theta_3)}{\sin(\theta_1)\sin(\theta_2)\sin(\theta_3)}
\end{equation}
4. $U_q(so(3,1)),q=e^{-\kappa},N=\frac{\pi}{2\kappa \sinh^2(\kappa)}.$
\begin{eqnarray}
\Theta(\rho_1,\rho_2,\rho_3)&=&\sum_{n=1}^{+\infty}\frac{1}{[n]_q}
\prod_{k=1}^3\frac{\sin(n\kappa \rho_{k})}{\sin(\kappa \rho_{k})}\\
&=&  \frac{1}{8} \frac{\vartheta_1(\kappa \rho_1)\vartheta_1(\kappa \rho_2)\vartheta_1(\kappa \rho_3)}{\sin(\kappa \rho_1) \sin(\kappa \rho_2) \sin(\kappa \rho_3)}\frac{q^{1/4}(q^{-1}-q)(q^2;q^2)^3_{\infty}}{\vartheta_4(\kappa\frac{\rho_1+\rho_2+\rho_3}{2})\vartheta_4(\kappa\frac{-\rho_1+\rho_2+\rho_3}{2})\vartheta_4(\kappa\frac{\rho_1-\rho_2+\rho_3}{2})\vartheta_4(\kappa\frac{\rho_1+\rho_2-\rho_3}{2})}\nonumber
\end{eqnarray}
where the Jacobi functions $\vartheta_4$ and $\vartheta_1$ are defined by:
\be
\vartheta_4(z) = \sum_{n=-\infty}^{+\infty} (-1)^n q^{n^2} \exp(2inz) \;\; , \;\; \vartheta_1(z)=-i\exp (iz  - \frac{\kappa}{4}) \vartheta_4(z + \frac{i\kappa}{2})\;.
\ee


\begin{thebibliography}{99}
\bibitem{Barrett}J.W.Barrett,
{\sl ``Geometrical measurements in three-dimensional quantum gravity'',}
Xth Oporto Meeting on Geometry, Topology and Physics, September 2001. {\tt [arXiv:gr-qc/0203018 ]}.

\bibitem{FL6j} L.Freidel, D.Louapre,
{\sl ``Asymptotics of 6j and 10j symbols''}, Class.Quant.Grav. 20 (2003) 1267-1294, {\tt [arXiv:hep-th/0209134] }.

\bibitem{NR}K.Noui, Ph.Roche,
{\sl ``Cosmological Deformation of Lorentzian Spin Foam Models'',}
Class.Quant.Grav. 20 (2003) 3175-3214, {\tt [arXiv:gr-qc/0211109 ]}.

\bibitem{BC}
  J.~W.~Barrett and L.~Crane,
  ``A Lorentzian signature model for quantum general relativity,''
  Class.\ Quant.\ Grav.\  {\bf 17}, 3101 (2000)
{\tt   [arXiv:gr-qc/9904025].}

\bibitem{FL}
L.Freidel and E.Livine,
{ ``Ponzano-Regge model revisited III: Feynman diagrams and Effective field theory''}
{\tt [arXiv: hep-th/0502106].} 

\bibitem{FK}
L.Freidel, K.Krasnov,
``Simple Spin Networks as Feynman Graphs,''
J.Math.Phys. 41 (2000) 1681-1690, {\tt ep-th/9903192 }


\bibitem{PRI}
L Freidel, D Louapre,
{\it Ponzano-Regge model revisited I: Gauge fixing observables and Lorentzian gravity},
Class.\ Quant.\ Grav.\  {\bf 21}, 5685 (2004),
{\tt [arXiv: hep-th/0401076.]}

\bibitem{BarrettFeynman}
  J.~W.~Barrett,
  ``Feynman diagams coupled to three-dimensional quantum gravity,''
 {\tt [ arXiv:gr-qc/0502048.]}

\bibitem{PRIII}
  L.~Freidel and E.~R.~Livine,
  ``Ponzano-Regge model revisited. III: Feynman diagrams and effective field
  theory,''
 {\tt [ arXiv:hep-th/0502106.]}

\bibitem{Noui}
K. Noui,
``Three dimensional Loop Quantum Gravity: particles and the quantum double'', {\tt [ submitted to Phys. Rev. D] };
K. Noui and A. Perez,
``Three dimensional Loop Quantum Gravity: coupling to point particles'',
Class. Quant. Grav. {\bf 22}, 4489, (2005).

\bibitem{CP}
V.Chari and A.Pressley, ``A guide to Quantum Groups'', Cambridge University press, Cambridge, 1994.

\bibitem{Ma}
S.Majid,``Foundations of Quantum Group Theory,'' Cambridge University Press (1995).

\bibitem{KM}
T.H.Koornwinder and N.M.Muller,``The quantum double of a locally compact group,''J.Lie Theory {\bf 7} (1997), 33-52 (1997).












\bibitem{BR1}E.Buffenoir, Ph.Roche,
{\sl ``Harmonic Analysis on the quantum Lorentz group'',}
Commun.Math.Phys. 207 (1999) 499-555, {\tt [arXiv:q-alg/9710022]}.

\bibitem{BR2}E.Buffenoir, Ph.Roche,
{\sl ``Tensor Products of Principal Unitary Representations of Quantum Lorentz
Group and Askey-Wilson Polynomials''}, J.Math.Phys. 41 (2000) 7715-7751, {\tt [arXiv: math.QA/9910147].}

\bibitem{DPR}R.Dijkgraaf, V.Pasquier, Ph.Roche,
{\sl ``Quasi Quantum Groups
and Orbifold Models''}.Nucl.Phys.Proc Suppl.{ 18B},
 60-72, (1990).

\bibitem{BarrettCrane}
  J.~W.~Barrett and L.~Crane,
  ``A Lorentzian signature model for quantum general relativity,''
  Class.\ Quant.\ Grav.\  {\bf 17}, 3101 (2000)
 {\tt  [arXiv:gr-qc/9904025].}

\bibitem{Baezbarrett}
  J.~C.~Baez and J.~W.~Barrett,
  ``Integrability for relativistic spin networks,''
  Class.\ Quant.\ Grav.\  {\bf 18}, 4683 (2001)
 {\tt  [arXiv:gr-qc/0101107].}





\end{thebibliography}
\end{document}